\documentclass[a4paper,openany,12pt]{article}
\usepackage[utf8]{inputenc} 
\usepackage{bbold}
\usepackage{dsfont}
\usepackage{amssymb}
\usepackage{enumerate}
\usepackage{anysize}
\usepackage{verbatim}
\usepackage{amsmath}
\usepackage{slashed}
\usepackage{cite}
\usepackage{tikz} 
\usetikzlibrary{shapes,arrows,positioning,automata,backgrounds,calc,er,patterns}
\usepackage{tikz-feynman}
\tikzfeynmanset{compat=1.0.0}
\usepackage[none]{hyphenat}
\usepackage{float}
\usepackage{feynmp-auto} 
\usepackage{setspace}
\marginsize{2.0 cm}{2.0 cm}{2.0 cm}{2.0 cm} 
\usepackage{subfigure}
\usepackage[normalem]{ulem}
\usepackage{cancel}
\usepackage{xcolor,colortbl}
\usepackage[colorlinks=true, urlcolor=blue, linkcolor=blue, citecolor=blue]{hyperref}
\usepackage{cleveref}
\usepackage{boldline}
\usepackage{makecell}

\definecolor{cosmiclatte}{rgb}{1.0, 0.97, 0.91}
\definecolor{lightcyan}{rgb}{0.88, 1.0, 1.0}

\newcommand*{\olra}{\overleftrightarrow}

\title{Effective field theory analysis of dark matter-standard model interactions with spin one mediators}
\date{}
\author{Fabiola Fortuna$^1$, Pablo Roig$^1$ and Jos\'e Wudka$^2$ \\
$^{1}$ \small{Centro de Investigaci\'on y de Estudios Avanzados,} \\
\small{Apartado Postal 14-740, 07000, Ciudad de M\'exico, M\'exico} \\
$^{2}$\small{Department of Physics and Astronomy, UC Riverside, Riverside, CA 92521, USA}
}

\begin{document}

\maketitle

\section*{Abstract}
We analyze interactions between dark matter and standard model particles with spin one mediators in an effective field theory framework. In this paper, we are considering dark particles masses in the range from a few MeV to the mass of the $Z$ boson. We use bounds from different experiments: $Z$ invisible decay width, relic density, direct detection experiments, and indirect detection limits from the search of gamma-ray emissions and positron fluxes. We obtain solutions corresponding to operators with antisymmetric tensor mediators that fulfill all those requirements within our approach. 

\section{Introduction}\label{sec:Intro}
All evidences of dark matter (DM) are purely gravitational \cite{Zwicky:1933gu, Rubin:1970zza, Rubin:1980zd, Corbelli:1999af, Clowe:2006eq, Allen:2011zs, Ade:2015xua} and -unfortunately- all direct \cite{Fu:2016ega, Aprile:2017iyp, Akerib:2016lao, Behnke:2016lsk, Akerib:2016vxi, Tan:2016zwf, Hanany:2019lle}, indirect \cite{Hooper:2010mq, Bulbul:2014sua, Urban:2014yda, Choi:2015ara, Ruchayskiy:2015onc, Ackermann:2015zua, Franse:2016dln, Aharonian:2016gzq, Cui:2016ppb, Aartsen:2016zhm, TheFermi-LAT:2017vmf} and collider searches \cite{Abercrombie:2015wmb, Aaboud:2016wna, Sirunyan:2018gka, ATLAS} for its detection have been unsuccessful so far.

Usually, though not always \cite{Baer:2014eja}, a weakly interacting massive particle (WIMP) \cite{Dutra,Roszkowski:2017nbc} is the preferred DM candidate. This is because, under this paradigm, the coupling of dark sector particles to the Standard Model (SM) is weak  enough to avoid the direct and indirect detection limits, yet sufficiently strong to generate a relic abundance complying with the one inferred from measurements of the cosmic microwave background radiation (which requires annihilation cross sections in the electroweak range). This is straightforwardly obtained in many popular new physics scenarios at the TeV scale. Generally, one obtains DM masses in the range from a few GeV to 1 TeV with the dark and SM sectors interacting through the exchange of mediators with electroweak-strength couplings to both sectors; although DM masses, three orders of magnitude smaller or larger can also be realized without fine-tuning the corresponding couplings \cite{Roszkowski:2017nbc}.

In absence of any direct DM signal, the generality of the effective field theory (EFT) approach may be advantageous (see e.g. refs. \cite{Belanger:2008sj, Goodman:2010ku, Crivellin:2014gpa, Crivellin:2014qxa, Duch:2014xda} in this context), as it only uses the known SM symmetries and degrees of freedom, assuming that the typical energy of all relevant processes lies below the mediator mass. Here we will continue exploring the phenomenological consequences of the EFT scenario developed in ref. \cite{Wudka} for the interactions between SM and DM particles with heavy mediators. In particular, we will focus on the case of spin-one mediators (either in the Proca  or antisymmetric tensor representations), that has received considerably less attention in the literature than the Higgs portal (see ref. \cite{Arcadi:2019lka} and  references therein) and neutrino portal \cite{Cosme:2005sb, An:2009vq, Falkowski:2009yz, Lindner:2010rr, Farzan:2011ck, Falkowski:2011xh, Heeck:2012bz, Baek:2013qwa, Baldes:2015lka, VaniaIllana, Batell:2017rol, HajiSadeghi:2017zrl, Bandyopadhyay:2018qcv, Berlin:2018ztp, Blennow:2019fhy} cases. According to the WIMP freeze-out scenario  motivated above, DM masses naturally lie in the GeV-TeV range. We will focus in this work in the low-energy region that is the one most restricted by observations, with DM masses under $M_Z$ (we take this upper bound instead of $M_H$ because in our EFT, considering only interactions with spin-one mediators, the invisible Higgs decay width is not modified at leading order). In this case, even the heavy mediators could eventually be found at the LHC.

For DM particle mass in this region the main observational/experimental constraints are: the invisible $Z$ decay width \cite{pdg} \footnote{Other $W, Z$ boson decays give less restrictive constraints.}, the observed relic density \cite{pdg} and direct detection limits from Xenon1T \cite{Aprile:2018dbl}, PandaX \cite{Ren:2018gyx}, LUX \cite{Akerib:2018hck}, DarkSide-50 \cite{Agnes:2018oej} and CRESST-III \cite{Abdelhameed:2019hmk}. We also employed recent indirect detection bounds (searching for excess gamma ray emissions) derived from dwarf spheroidal galaxy observations released by the Fermi-LAT and DES Collaborations \cite{Drlica}. We also used the results of indirect DM searches based on antimatter detection, specifically, the limits on the annihilation cross section derived in ref. \cite{Ibarra:2013zia} using the AMS-02 data on the positron flux \cite{Accardo:2014lma}.

This paper is organized as follows: in section \ref{sec:EFT} we introduce the effective field theory that we are using \cite{Wudka} and highlight the part interesting for our study and our conventions. Then, in section \ref{sec:GammaZ} we compute the bounds that the invisible $Z$ decay width puts on our several possible DM candidates. After that, in section \ref{sec:RelicDensity} we verify that the observed relic density can be reproduced in the different cases. Next, in section \ref{sec:ObsLims} we analyze some observational limits: in subsection \ref{subsec:DirectDetection} we check that direct detection bounds are respected; and in subsections \ref{subsec:DSSG} and \ref{subsec:AMS02} we consider the indirect bounds given by dwarf spheroidal galaxies and the positron flux, respectively. We discuss our results and conclude in section \ref{sec:Concl}.

\section{Effective field theory}\label{sec:EFT}
A simple way to ensure the dark sector contains a stable particle that will play the role of DM is the following: we will assume \cite{darkmatter, Feng:2010gw} that all dark fields transform non-trivially under a symmetry group $\mathcal{G}_\text{DM}$ (whose nature we will not need to specify), while all SM particles are hypothesized to be $\mathcal{G}_\text{DM}$ singlets. Also, we assume that all dark fields are singlets under the SM gauge group $\mathcal{G}_\text{SM}=SU(3)\otimes SU(2)\otimes U(1)$.

Effective field theory (EFT) formulations of dark matter interactions have proven to be a convenient and popular way to quantify many bounds on dark matter \cite{Goodman:2010ku,Duch:2014xda,Racco:2015dxa,Bell:2015sza,DeSimone:2016fbz,Cao:2009uw,Cheung:2012gi,Busoni:2013lha,Buchmueller:2014yoa}.
We will follow this approach, assuming that the SM-DM interactions are generated by the exchange of heavy mediators that we take to be singlets under $\mathcal{G}_\text{DM} \times \mathcal{G}_\text{SM}$. The interactions between the dark and Standard-Model sectors then take the form
\begin{equation} \label{ops}
	\mathcal{O} = \mathcal{O}_\text{SM} \mathcal{O}_\text{dark}\,,
\end{equation} 
where $\mathcal{O}_\text{SM,dark}$ denote local operators composed of standard model and dark matter fields, respectively (singlets under  $\mathcal{G}_\text{SM,DM}$, respectively).

When constructing $\mathcal{O}_\text{dark}$ we will assume \cite{Wudka} that the dark sector is composed of scalars $\Phi$, Dirac fermions $\Psi$ and Proca vectors $X$, with the understanding that the dark sector present in Nature may only contain a subset of these particles.

Since all dark fields are assumed to transform non-trivially under $\mathcal{G}_\text{DM}$, $\mathcal{O}_\text{dark}$ will contain at least two fields. The list of the operators $\mathcal{O}$ of dimension $\leq 6$ satisfying the above conditions is given in table \ref{tab:effops} \cite{Wudka}.

\renewcommand{\arraystretch}{1.7}

\begin{table}[h!]
  \begin{center}
    \begin{tabular}{c|c|c} 
      \rowcolor{lightcyan}
      \textbf{dim.} & \textbf{category} & \textbf{operator(s)}\\
      \Xhline{1.5pt}
      4 & I & $ \vert \varphi \vert^2 (\Phi^\dagger \Phi) $  \\
      \Xhline{1.2pt}
      \rowcolor{cosmiclatte}      
      & II & $ \vert \varphi \vert^2 \bar{\Psi} \Psi \hspace{1cm} \vert \varphi \vert^2 \Phi^3 $\\
      5 & III & $ (\bar{\Psi} \Phi) (\varphi^T \epsilon \ell) $ \\
	  \rowcolor{cosmiclatte}            
      & IV & $ B_{\mu \nu} X^{\mu \nu} \Phi \hspace{1cm} B_{\mu \nu} \bar{\Psi} \sigma^{\mu \nu} \Psi $ \\
      \Xhline{1.2pt}
      & V & $ \vert \varphi \vert^2 \mathcal{O}^{(4)}_{\text{dark}} \hspace{1cm} \Phi^2 \mathcal{O}^{(4)}_{\text{SM}}  $ \\
	  \rowcolor{cosmiclatte}            
      6 & VI & $ (\bar{\Psi} \Phi^2) (\varphi^T \epsilon \ell) \hspace{1cm} (\bar{\Psi} \Phi) \slashed{\partial} (\varphi^T \epsilon \ell) $ \\
      & VII & $ \mathcal{J}_{\text{SM}}^\mu \mathcal{J}_{\text{dark}\,\mu} $ \\
	  \rowcolor{cosmiclatte}            
      & VIII & $ B_{\mu \nu} \mathcal{O}^{(4) \mu \nu}_{\text{dark}} $ \\     
    \end{tabular}
  \end{center}
  \caption{Effective operator list up to dimension $6$ involving dark and SM fields;  $\varphi$ stands for the SM scalar isodoublet, $B$ for the hypercharge gauge field, and $\ell$ is a left-handed lepton isodoublet; also, $\epsilon = i \sigma_2$, where $\sigma_2$ is the corresponding Pauli matrix. Dark scalars, Dirac dark fermions and vectors are denoted by $\Phi$, $\Psi$ and X, respectively. The vector currents operators in category VII are defined in \cref{corrientes}, and the operators $\mathcal{O}^{(4)}_{\text{dark} \hspace{1mm} \mu \nu}$ in category VIII are given 
in \cref{opDM2}.} \label{tab:effops}      
\end{table}

The $\mathcal{O}^{(4)}$ of categories V and VIII represent dimension 4 local operator combinations of the corresponding sector; the relevant ones for this work are: 
\begin{equation} \label{opDM2}
	\mathcal{O}^{(4)}_{\text{dark} \hspace{1mm} \mu \nu} \in \lbrace \Phi^\dagger X_{\mu \nu} \Phi, \Phi \bar{\Psi} \sigma_{\mu \nu} P_{L,R} \Psi, \bar{\Psi} (\gamma_\mu \overleftrightarrow{\mathcal{D}}_\nu - \gamma_\nu \overleftrightarrow{\mathcal{D}}_\mu) P_{L, R} \Psi \rbrace.      
\end{equation}

In category VII, $\mathcal{J}_\text{SM, dark}^\mu$ represents a dimension 3 vector current, either dark or standard:  
\begin{equation} \label{corrientes}
\begin{array}{cc} 
	\mathcal{J}_\text{SM}^{(\psi) \mu} = \bar{\psi} \gamma^\mu \psi, & \mathcal{J}_\text{SM}^{(\varphi) \mu} = \frac{1}{2i} \varphi^\dagger \overleftrightarrow{D}^\mu \varphi,  \\
   \mathcal{J}_\text{dark}^{(L,R) \mu} = \bar{\Psi} \gamma^\mu P_{L,R} \Psi, & \mathcal{J}_\text{dark}^{(\Phi) \mu} = \frac{1}{2i} \Phi^\dagger \overleftrightarrow{\mathcal{D}}^\mu \Phi,
\end{array}
\end{equation}
where $\psi$ denotes any SM fermion, $D$ the covariant derivative in the standard sector, and $\mathcal{D}$ the covariant derivative in the dark sector (replaced by an ordinary derivative if this sector is not gauged).

We will assume that the full theory -- composed of mediators, dark and standard sectors -- is renormalizable. Within the neutral-mediator paradigm one can determine by inspection \cite{Wudka} that the operators in table \ref{tab:effops} are generated at tree level \footnote{In category I, there is the Higgs-portal, renormalizable operator.} by scalar mediators (categories II an V), fermion mediators (categories III and VI), vector mediators (category VII), or antisymmetric tensors representing spin-one mediators (categories IV and VIII). In this work, we will focus on operators with vector and antisymmetric tensor mediators because the models with scalar and fermion mediators have already been studied extensively \cite{Goodman:2010ku,Duch:2014xda,Racco:2015dxa,Bell:2015sza,DeSimone:2016fbz,Cao:2009uw,Cheung:2012gi,Busoni:2013lha,Buchmueller:2014yoa,Patt:2006fw,VaniaIllana,Lamprea19}. We emphasize that the  type of vector mediators we refer to do not correspond to dark photons \cite{Okun:1982xi, Holdom:1985ag}, which in the current scheme are members of the dark sector transforming non-trivially under ${\cal G}_{\rm dark}$, while the vector mediators considered here are singlets under this group.

Discussing the detailed UV completion of this effective field theory is beyond the scope of this work. Such completions have been discussed in the literature for the case of spin zero mediators \cite{Baek:2011aa,LopezHonorez:2012kv,Berlin:2015wwa}, fermion mediators \cite{Bai:2013iqa} and vector mediators associated with a gauged $B-L$ current (see {\it e.g.} ref. \cite{Klasen:2016qux} and references therein). The antisymmetric tensor mediator case\footnote{In principle the operators obtained integrating these fields out do not need to come from loop-level processes (see e. g. refs. \cite{Kalb:1974yc, Rohm:1985jv} for their appearance in string theory).} can be problematic because renormalizable theories require such fields to be coupled to a conserved 2-index current of dimension $ \le3 $, which the SM does not possess. This issue can be addressed by including additional degrees of freedom at higher energy scales \cite{Cata:2014sta}.

The Lagrangian we use is conveniently separated into two parts:
\begin{itemize} 
\item Terms involving dark fermions ($\Psi$):
\begin{equation} \label{dfermion}
	\mathcal{L}_\text{eff}^\Psi = \frac{\Upsilon_\text{eff}}{\Lambda} B_{\mu \nu} \bar{\Psi} \sigma^{\mu \nu} \Psi + \frac{A_\text{eff}^{L,R}}{\Lambda^2} \bar{\psi} \gamma_\mu \psi \bar{\Psi} \gamma^\mu P_{L,R} \Psi + \frac{\kappa_\text{eff}^{L,R}}{\Lambda^2} B_{\mu \nu} \bar{\Psi} (\gamma^\mu \protect\olra{\mathcal{D}}^\nu - \gamma^\nu \protect\olra{\mathcal{D}}^\mu) P_{L, R} \Psi.
\end{equation}

\item Terms involving dark bosons ($X,\,\Phi$):  
\begin{equation} \label{dsdv}
	\mathcal{L}_\text{eff}^{\Phi,X} = \frac{\zeta_\text{eff}}{\Lambda} B_{\mu \nu} X^{\mu \nu} \Phi + \frac{\epsilon_\text{eff}}{\Lambda^2} \bar{\psi} \gamma_\mu \psi \frac{1}{2i} \Phi^\dagger \overleftrightarrow{\mathcal{D}}^\mu \Phi.
\end{equation}
\end{itemize}

In the calculations below we generally adopt the single operator dominance hypothesis, for example when computing $\Gamma_{Z \rightarrow \bar{\Psi} \Psi}$, we first take $\Upsilon_\text{eff} \neq 0$ and $\kappa^{L,R}_\text{eff}=0$ and then the opposite. We comment of the combined effects of some operators in Sect. \ref{sec:Concl}.

\section{Invisible $Z$ decay width}\label{sec:GammaZ}

As noted in section \ref{sec:Intro} we consider DM masses below $M_Z$, in which case the invisible decay of the $Z$ places important constraints on the parameters of our model related to operators in categories IV and VIII of table \ref{tab:effops}.

Recently, ref. \cite{Janot:2019oyi} improved the prediction of Bhabha scattering $e^+e^- \rightarrow e^+e^-$, and when accounted for, the corresponding changes modify the number of light neutrino species -- as determined from LEP measurement of the hadronic cross section at the $Z$ peak; the new value being $N_\nu = (2.9975 \pm 0.0074)$. Using these results the experimental value of the $Z$ invisible decay width becomes $\Gamma^\text{inv}_Z = (501.03 \pm 1.27)$ MeV, which includes the standard decays to neutrino-antineutrino pairs and may include the decay to any other undetected contributions. The theoretical value (SM) of the partial decay rate to a pair of neutrinos is \mbox{$\Gamma(Z \rightarrow \bar{\nu} \nu) = (167.15 \pm 0.01)$ MeV} \cite{pdg}. Assuming the existence of three light active neutrinos, this yields $\Gamma^\text{inv}_Z - \Gamma^{\bar{\nu} \nu}_Z = (-0.42 \pm 1.30)$ MeV. We will then use \footnote{\label{ft_iv} Employing instead the value $\Gamma^\text{inv}_Z = (499.0\,\pm\,1.5)$ MeV reported in the PDG \cite{pdg}, we  obtain \mbox{$\Gamma^\text{inv}_Z - \Gamma^{\bar{\nu} \nu}_Z = 0.49\; \text{MeV} \hspace{2mm} \text{at} \hspace{2mm} 95 \% \text{CL}$.}}   
\begin{equation} \label{Zinv}
	\Gamma^\text{inv}_Z - \Gamma^{\bar{\nu} \nu}_Z = 2.13\; \text{MeV} \hspace{2mm} \text{at} \hspace{2mm} 95 \% \text{CL}.      
\end{equation}
  	
\subsection{Dark fermions} \label{sub:darkfermions}

Using the operator $B_{\mu \nu} \bar{\Psi} \sigma^{\mu \nu} \Psi$ in \cref{dfermion} to calculate $Z \rightarrow \bar{\Psi} \Psi$ we get the matrix element squared
\begin{equation}
	\overline{ \vert \mathcal{M} \vert^2}  = \frac{8 \Upsilon_\text{eff}^2   \sin^2 \theta_W}{3 \Lambda^2} m_Z^2 (8 m_\Psi^2 + m_Z^2), 
\end{equation}
that we use to write the effective Lagrangian coefficient in terms of the partial decay width:
\begin{equation}
	\frac{\Upsilon_\text{eff}}{\Lambda} = \left\lbrace \frac{6 \pi \Gamma_{Z \rightarrow \bar{\Psi} \Psi}}{  \sin^2 \theta_W (8m_\Psi^2 + m_Z^2)\sqrt{m_Z^2 - 4 m_\Psi^2} } \right\rbrace^\frac{1}{2}.
\end{equation}

\begin{figure}[h] 
\centering
\includegraphics[width=105mm]{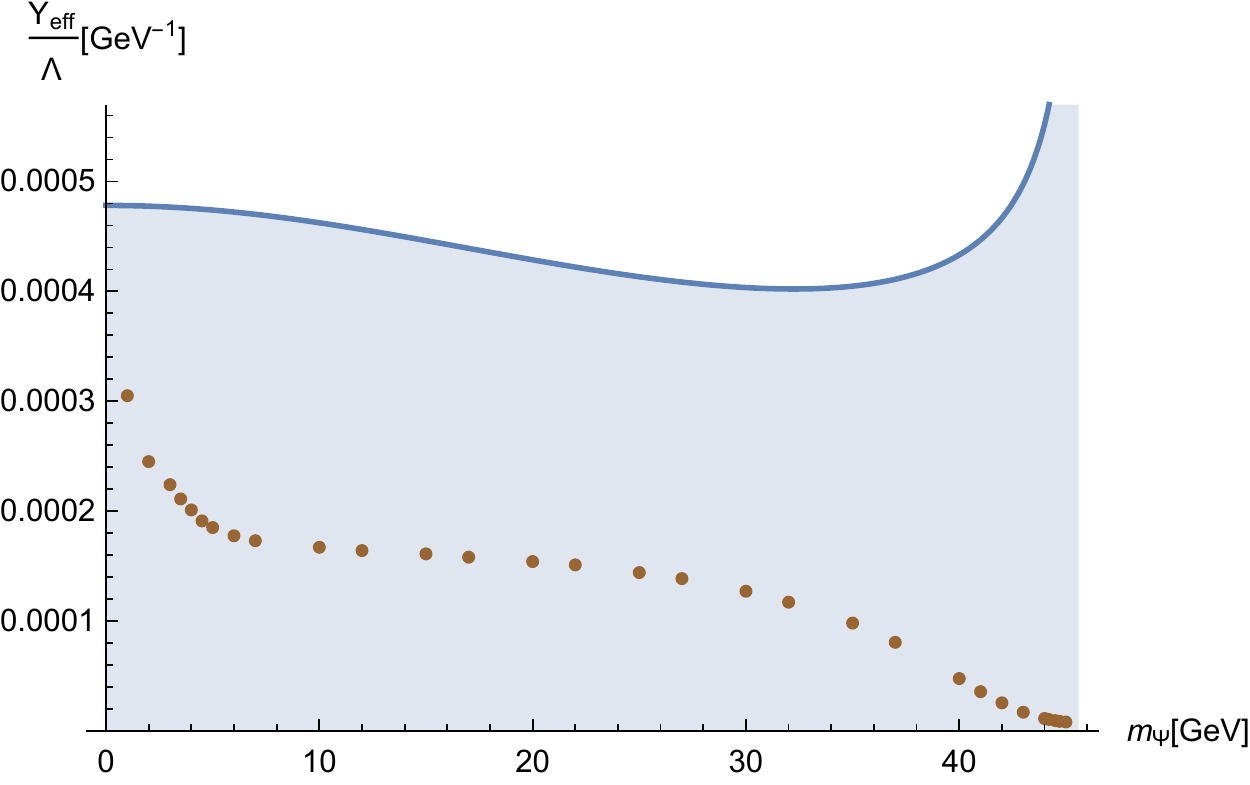}
\caption{Effective coupling $\Upsilon_\text{eff}/\Lambda$ as a function of the dark fermion mass, compatible with the invisible decay width of the $Z$ boson (blue area) and the observed relic density (brown dots)---see section \ref{sec:RelicDensity} for further detail---.} \label{fig:DRcom1}
\end{figure}

Using \cref{Zinv} for $\Gamma_{Z \rightarrow \bar{\Psi} \Psi}$ and $m_Z=91.1876(21) \hspace{1mm} \text{GeV}$ \cite{pdg}, we plot in fig. \ref{fig:DRcom1} the region allowed by this constraint in the $ m_\Psi - \Upsilon_\text{eff}/\Lambda $ plane (shaded blue area). 

If we use the operator $B^{\mu \nu} \bar{\Psi} (\gamma_\mu \overleftrightarrow{\mathcal{D}}_\nu - \gamma_\nu \overleftrightarrow{\mathcal{D}}_\mu) P_{L, R} \Psi$ 
we get the following expression
\begin{equation} \small
\Gamma_{Z\to\bar\Psi\Psi} = \frac{  \sin^2 \theta_W m_Z^2 \sqrt{m_Z^2 - 4 m_\Psi^2}}{6 \pi \Lambda^4} \left[ m_Z^2 \left\lbrace \frac{}{} (\kappa_\text{eff}^L)^2 + (\kappa_\text{eff}^R)^2 \right\rbrace - m_\Psi^2 \left\lbrace \left(\kappa_\text{eff}^L \right)^2 -6 \kappa_\text{eff}^L \kappa_\text{eff}^R +\left(\kappa_\text{eff}^R \right)^2 \frac{}{} \right\rbrace \right].
\end{equation}

\begin{figure}[h] 
\centering
\includegraphics[width=105mm]{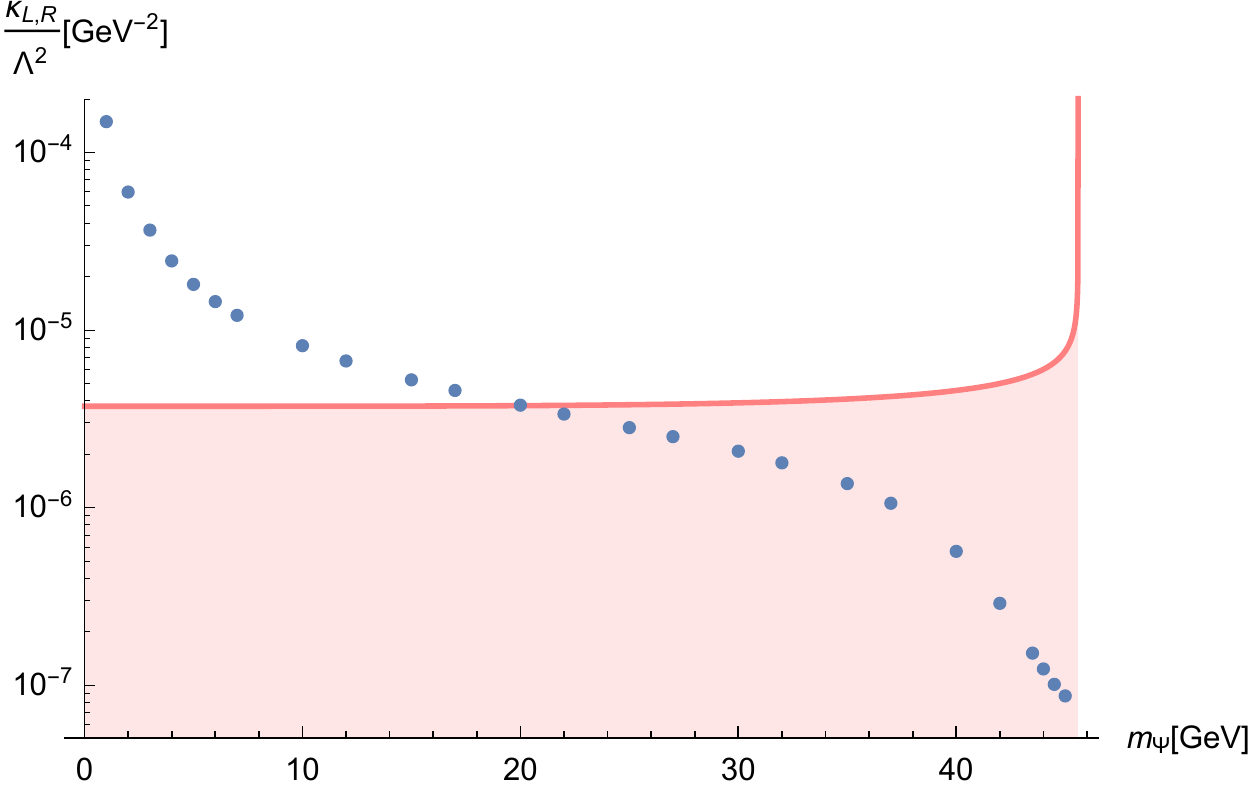}
\caption{$\kappa_\text{eff}^{L,R}/\Lambda^2$ as a function of the dark fermion mass, compatible with the invisible decay width of the $Z$ boson (pink area) and the observed relic density (blue dots)---see section \ref{sec:RelicDensity} for further detail---.} \label{fig:DRcom3}
\end{figure}

We take $\kappa_\text{eff}^L = \kappa_\text{eff}^R$ in order to reduce the number of free parameters, and we obtain
\begin{equation} \label{ZinvOp6}
\frac{\kappa_\text{eff}^{L,R}}{\Lambda^2} = \left\lbrace \frac{3 \pi \Gamma_{Z \rightarrow \bar{\Psi} \Psi}}{  \sin^2 \theta_W m_Z^2 \sqrt{m_Z^2 - 4 m_\Psi^2} (2m_\Psi^2 + m_Z^2)} \right\rbrace^\frac{1}{2},
\end{equation}  
from which we obtain the allowed shaded (pink) region in fig. \ref{fig:DRcom3}.

Next we select benchmark values for the effective operator coefficients and values of $ \Lambda $. To motivate our choices consider, for example, the first term in \cref{dfermion}, generated at tree-level by the exchange of a heavy antisymmetric tensor field, and for which we estimate the scale $\Lambda$ using fig. \ref{fig:DRcom1}:
\begin{equation} \label{cota}
	\frac{\Upsilon_\text{eff}}{\Lambda} \sim 
	4.4 \times 10^{-4} \text{GeV}^{-1},
\end{equation}
that in the fundamental theory would have the form \footnote{A $1/\Lambda^2$ dependence is expected from the propagator of the heavy mediator, with mass of order $\Lambda$. The coupling of the mediator to the $B$ has a coefficient with dimensions of mass, which we assume to be $ g_2 \Lambda  $.} 
\begin{equation} \label{relfun}
	\frac{\Upsilon_\text{eff}}{\Lambda}  = \frac{g_1 g_2}{\Lambda}
\end{equation}
where $g_1$, $g_2$ are the couplings in each vertex. 
We take the values of $g_1$, $g_2$ in the interval between that of the electron charge $ g_{1,2} \sim 0.3 $ and the weak coupling $g_{1,2} \sim 0.66$, as our educated guess. Using \cref{cota} we then find\footnote{Using the numbers in footnote \ref{ft_iv} we get instead $\Lambda\in[0.5,2.2]$ TeV.}
\begin{equation}
230 \text{GeV} < \Lambda < 1 \text{TeV}.
\label{escalas}
\end{equation}

Later we will use these numbers as reference values, also when we combine contributions of different operators, and to determine whether it is sufficient to work with the dimension 5 operators or if the operators of dimension 6 need to be considered as well. The estimated values of $\Lambda$ represent the largest energy scale up to which our EFT can possibly be used. Since the relevant scale for the processes that we consider is $E\lesssim M_Z$, the ratio $E/\Lambda$ is small enough to disregard operators of dimension $\geq7$.

\subsection{Dark Vector and Dark Scalar}

Using the operator $B_{\mu \nu} X^{\mu \nu} \Phi$ (category IV) in \cref{dsdv}, we compute $Z \rightarrow X \Phi$, and obtain
\begin{equation}
	\vert \mathcal{M} \vert^2 = \frac{4 \zeta^2_\text{eff} \sin^2 \theta_W}{3 M^2_\Omega} \left\lbrace \frac{1}{2} (m_Z^2 - m_\Phi^2 + m_X^2)^2 + m_X^2 m_Z^2 \right\rbrace,
\end{equation}
and the decay rate
\begin{equation}
 	\Gamma = \frac{\zeta^2_\text{eff} \, \sin^2 \theta_W}{12 \pi m_Z^3 \Lambda^2} \lambda^{\frac{1}{2}} (m^2_Z, m^2_\Phi, m^2_X) \left\lbrace \frac{1}{2} (m_Z^2 - m_\Phi^2 + m_X^2)^2 + m_X^2 m_Z^2 \right\rbrace,
\end{equation}
where $ \lambda(a,b,c) = a^2 + b^2  + c^2 - 2 a b - 2 b c - 2 c a $. Whence
\begin{equation} \label{exp}
	\frac{\zeta_\text{eff}}{\Lambda} = \left\lbrace \frac{12 \pi m_Z^3 \Gamma_{Z \rightarrow X \Phi}}{  \sin^2 \theta_W \lambda^\frac{1}{2} (m_Z^2, m_\Phi^2, m_X^2) \left[ \frac{1}{2} (m_Z^2-m_\Phi^2+m_X^2) + m_X^2 m_Z^2 \right]} \right\rbrace^\frac{1}{2}.
\end{equation}
As we did before, we use \cref{Zinv} to constrain the ratio $\zeta_\text{eff}/\Lambda$.

\begin{figure}[h] 
\centering
\includegraphics[width=105mm]{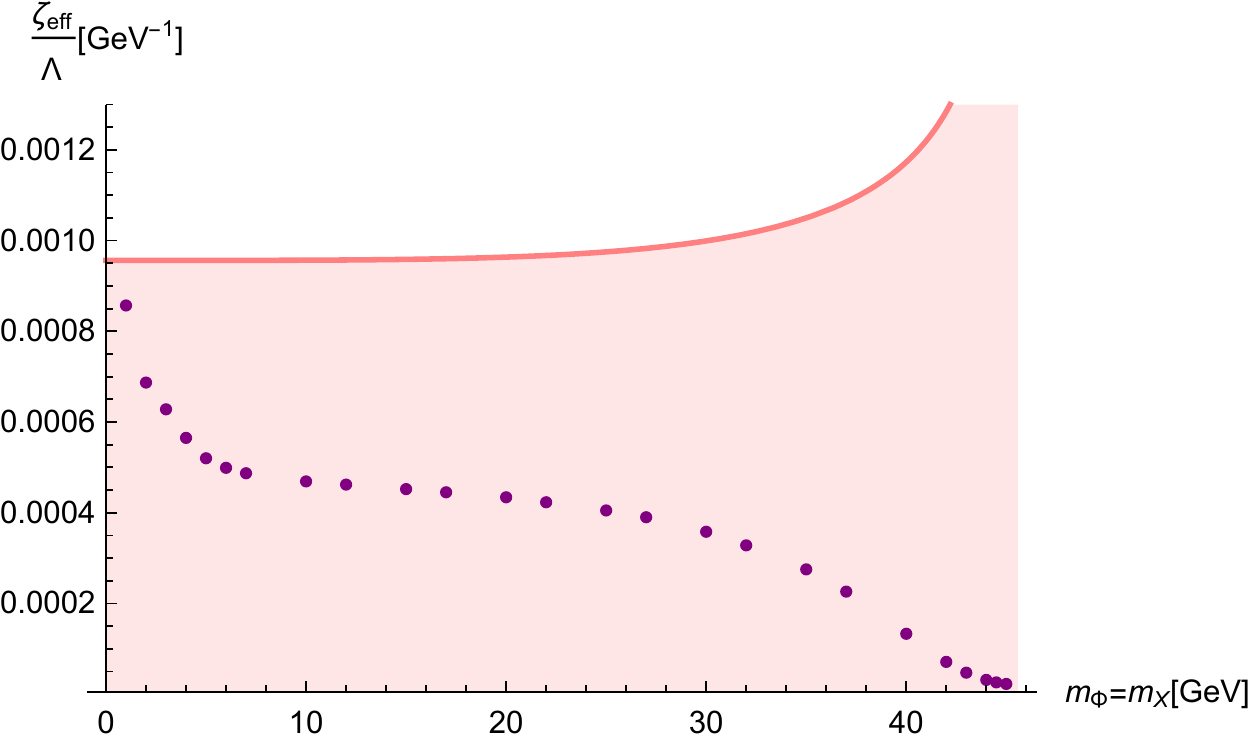}
\caption{Effective coupling $\zeta_\text{eff}/\Lambda$ as a function of the dark particles mass, compatible with the invisible decay rate of the $Z$ boson (pink area) and the observed relic density (violet dots)---see section \ref{sec:RelicDensity} for further detail---.} \label{fig:DRXScom}
\end{figure}

The above expression is a function of the masses of the vector and the scalar dark particles, but we will explain later that the only interesting values are those when $m_\Phi \sim m_X$. In this case, we obtain the allowed pink region in fig. \ref{fig:DRXScom}.

\section{Relic Density}\label{sec:RelicDensity}

We use micrOMEGAs code \cite{micromegas} to compute the relic abundance of dark matter in our EFT. We use just one operator at a time, and we obtain the coefficient~\footnote{Since the uncertainty in \cref{rd} is small, this constraint fixes the operator coefficient for each choice of the mass of the DM candidate(s).} in the Lagrangian --- in \mbox{eqs. \ref{dfermion} and \ref{dsdv}} --- such that they reproduce the observed relic density \cite{pdg19}  
\begin{equation} \label{rd}
	\Omega_\text{DM} h^2 = 0.1193 \pm 0.0009.
\end{equation}

\subsection{Dark fermions}
\begin{figure}[h] 
\centering
\includegraphics[width=105mm]{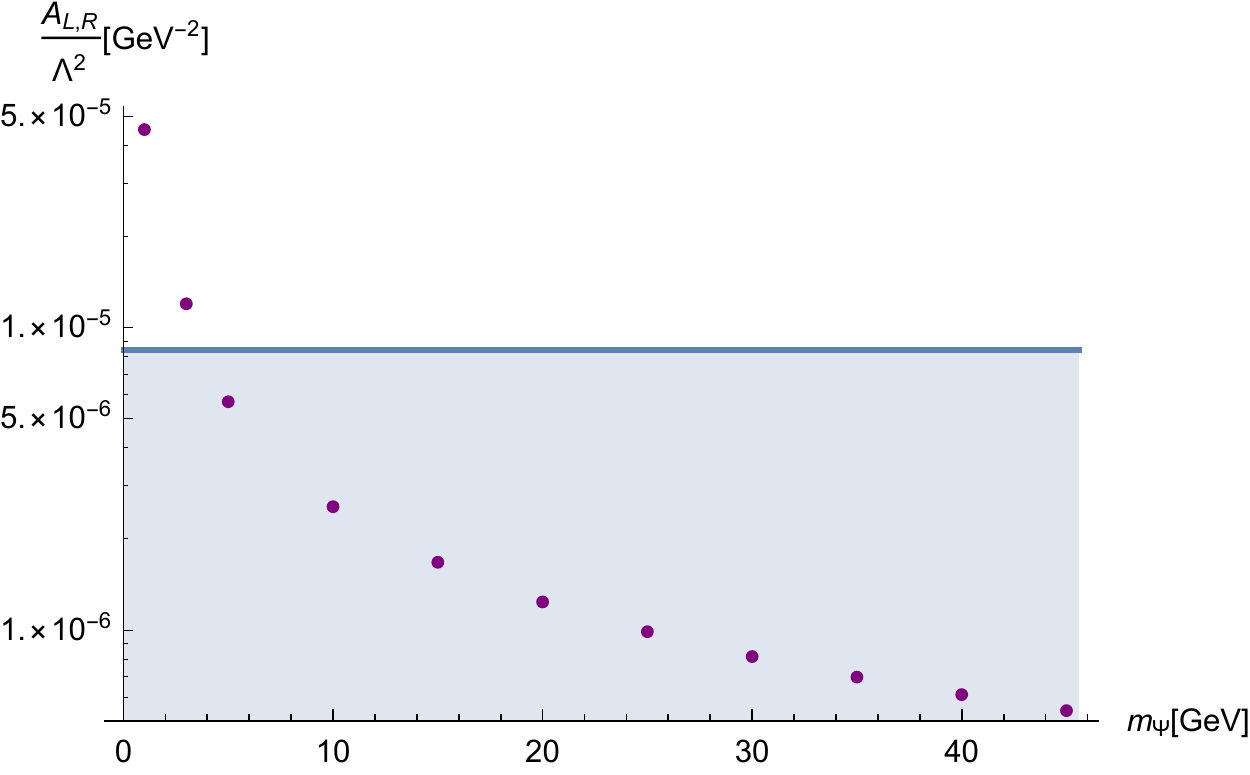}
\caption{Effective coupling $\frac{A_\text{eff}^{L, R}}{\Lambda^2}$ as a function of the dark fermion mass, compatible with the observed relic density (purple dots) and compared to a reference value, \cref{relajado}, shown by the blue line.} \label{fig:DRcom2}
\end{figure}

Proceeding as described above, and considering only the dimension 5 operator $ \propto \Upsilon_\text{eff}$ in \cref{dfermion}, we get the dots in fig. \ref{fig:DRcom1}. We can see that we have solutions compatible with both observations in all the mass range  we are analyzing. If we now use the operator with two fermionic currents, $\mathcal{J}_\text{SM}^{(\psi) \mu} \mathcal{J}_{\text{dark} \hspace{2mm \mu}}^{(L,R)}$ (category VII), we obtain the dots in fig. \ref{fig:DRcom2}, where we assume $A_\text{eff}^L=A_\text{eff}^R$. Finally, the computation of the relic density considering the operator with derivative couplings
$ \propto \kappa^{L,R}_\text{eff}$ in \cref{dfermion} (category VIII) generates the dots in fig. \ref{fig:DRcom3}, where we take $\kappa_\text{eff}^L=\kappa_\text{eff}^R$. In this case, we find solutions when $m_\Psi \gtrsim 20$ GeV. The case where we have both a dark scalar and a dark vector is discussed in the next subsection.

The couplings of the operators not contributing to the invisible decay of the $Z$ can be estimated as in \cref{relfun}
\begin{equation}
\begin{array}{|c|c|c|}
\hline
\rowcolor{lightcyan}
& g_{1,2} \sim e,\,\Lambda \sim 1\text{TeV} & g_{1,2} \sim 0.66,\,\Lambda \sim 230 \text{GeV} \cr \hline
g_1 g_2/\Lambda^2 \, (\text{GeV}^{-2}) & \sim 1.1 \times 10^{-7} & 8.4 \times 10^{-6} \cr \hline
\end{array}
\label{relajado}
\end{equation}

The values $A_\text{eff}^{L,R}$ consistent with the above estimates can be obtained from fig. \ref{fig:DRcom2}, where the constant value showed by the blue line corresponds to the least restrictive condition in \cref{relajado} and we observe that there are acceptable solutions when the fermion mass $m_\Psi \gtrsim 4$ GeV; we note that this corresponds to a ``small'' scale $\Lambda$ and/or ``large'' couplings $g_{1,2}$.

\subsection{Dark Vector and Dark Scalar} \label{sub:RDXS}

Turning next to the effects of the category IV operator of dimension 5, $ \propto \zeta_\text{eff}$ in \cref{dsdv}, we obtain the dots in fig. \ref{fig:DRXScom} where we use $m_X=m_\Phi$; in this case we find suitable solutions for the full range of masses that we are considering. 
In contrast, when $m_X$ and $m_\Phi$ differ by a few GeV or more, the model does not fit the constraints. This happens because, when $ m_X \sim m_\Phi $, the dominant process regulating the relic abundance is $ X \Phi \to \bar f f $ via an s-channel $B$ exchange; the cross section and invisible width are then $ \propto 1/\Lambda^4$. In contrast, when $ m_X > m_\Phi $ ($ m_\Phi > m_X $) the dominant process is $ \Phi\Phi \to \gamma\gamma $ ($XX \to \gamma\gamma $) via a t-channel $X$ ($\Phi$) exchange, which is quadratic in the effective vertex, giving an annihilation cross section $ \propto 1/\Lambda^8 $ (the invisible width is still $ \propto1/\Lambda^4$), relic abundance constraint then requires a value of $ \Lambda $ too small to be consistent with the $Z$ width.

\subsection{Dark Scalar}

\begin{figure}[h] 
\centering
\includegraphics[width=105mm]{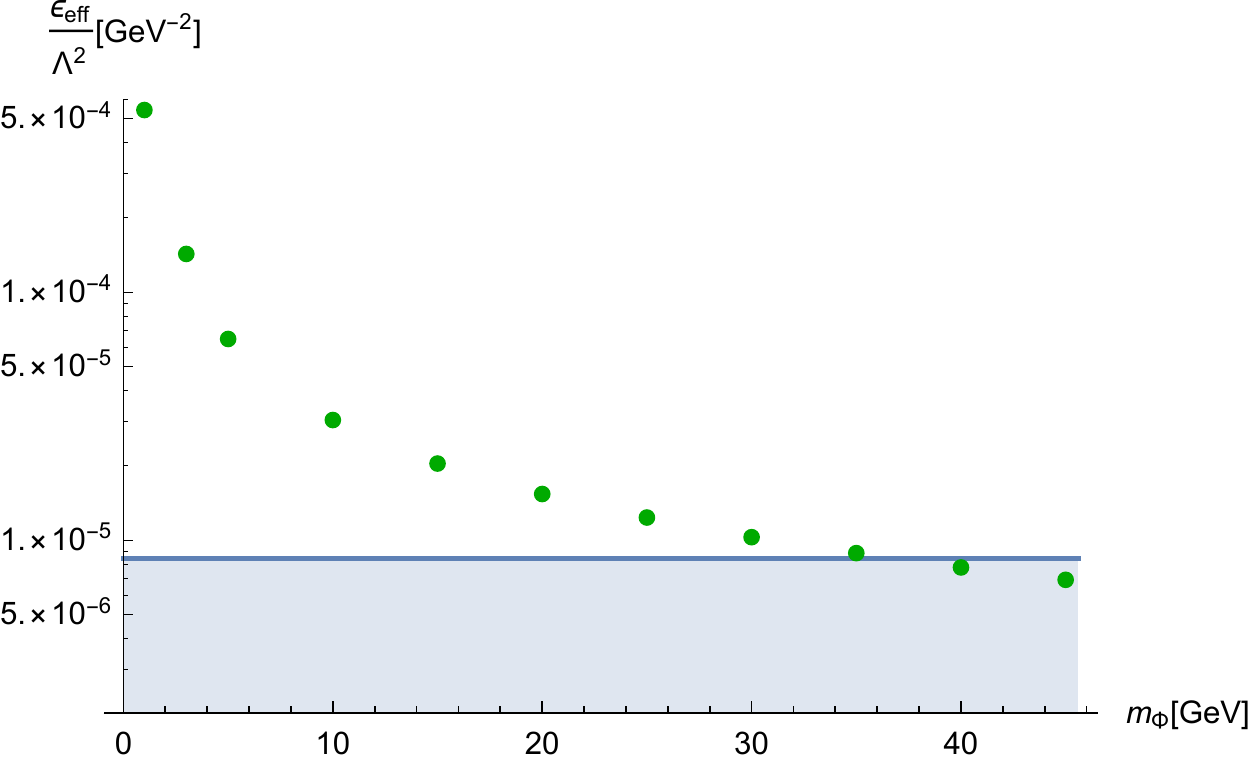}
\caption{Effective coupling $\frac{\epsilon_\text{eff}}{\Lambda^2}$ as a function of the dark scalar mass, compatible with the observed relic density (green dots) and compared to the reference value \cref{relajado}, shown by the blue line.} \label{fig:DRSScom}
\end{figure}

In this case the relevant operator (proportional to $ \epsilon_\text{eff}$ in \cref{dsdv}) is composed of a standard fermionic current and a dark scalar current. Using this we obtain the dots in fig. \ref{fig:DRSScom}, where the horizontal line corresponds to the weaker estimate in \cref{relajado}. Combining these results we find that only masses $ m_\Phi \gtrsim 36$ GeV are consistent with the benchmark range in \cref{relajado} .
     
\subsection{Other operators}

The remaining operators generated by spin one mediators differ from the ones treated so far in the number of dark particles in the effective vertex. In category VIII we have $B^{\mu \nu} \Phi^\dagger X_{\mu \nu} \Phi$, which contains 3 dark particles. In this case the invisible $Z$ width constraint is easily met because of the 3-body final state phase space suppression. The relic abundance (for $ m_\Phi \sim m_X $) is controlled by processes such as $ \Phi \Phi \to X \psi \bar\psi $ and $ \Phi X \to \Phi \psi \bar\psi $ with a virtual $B$ coupling to the standard fermions $ \psi $; these rates are also phase-space suppressed requiring an unreasonable small $ \Lambda $ in order to meet the experimental observations. The same issue arises with operator $B^{\mu \nu} \Phi \bar{\Psi} \sigma_{\mu \nu} P_{L,R} \Psi$. This situation is not a problem because, given a specific choice of DM particle(s), these dimension 6 operators  play a subdominant role (see the discussion at the end of subsection \ref{sub:darkfermions}). Inspection of figs. \ref{fig:DRcom2} and \ref{fig:DRSScom} reveals that suitable solutions can only be obtained for the largest \mbox{$\Lambda\sim 1$ TeV}. Because of this, we will neglect these subleading operators in our subsequent analysis.\footnote{Available freeware such as micrOMEGAs often assume a discrete symmetry within the dark sector to ensure DM stability. This excludes these dimension 6 operators, and makes it difficult to calculate their effects in detail.}

\section{Observational limits} \label{sec:ObsLims}

In this section we discuss limits derived from several direct and indirect detection constraints; we will use the following notation: 
\begin{align} \label{OPS}
	\text{OP1} &\equiv B_{\mu \nu} \bar{\Psi} \sigma^{\mu \nu} \Psi, \cr
	\text{OP2} &\equiv \bar{\psi} \gamma^\mu \psi \bar{\Psi} \gamma_\mu P_{L,R} \Psi, \cr
	\text{OP3} &\equiv B_{\mu \nu} \bar{\Psi} (\gamma^\mu \protect\olra{\mathcal{D}}^\nu - \gamma^\nu \protect\olra{\mathcal{D}}^\mu) P_{L, R} \Psi, \cr
	\text{OP4} &\equiv B_{\mu \nu} X^{\mu \nu} \Phi, \cr
	\text{OP5} &\equiv \frac{1}{2i} \left( \bar{\psi} \gamma^\mu \psi\right) \left( \Phi^\dagger \overleftrightarrow{\partial}_\mu \Phi\right).
\end{align}

In the calculations below we will use the effective couplings that correctly reproduce the relic density, \cref{rd}, as shown in the figures of sections \ref{sec:GammaZ} and \ref{sec:RelicDensity}. For the case of OP4 we only consider the scenario where $m_X = m_\Phi$, as discussed in subsection \ref{sub:RDXS}. We also consider the combined contributions from dimension 5 and 6 operators when they contain the same DM candidate; in such cases we adopt the following relationship between the scales $ \Lambda $ and operator coefficients $C$:
\begin{equation}
\Lambda_\text{dim 6} = \Lambda_\text{dim 5} \,, \qquad 	 C_\text{dim6} = \pm  C_\text{dim5},;
	\label{eq:combine}
\end{equation}
In these cases we find that the effects of the sign are negligible. We will again use the two previously estimated values for $\Lambda$, in \cref{escalas}, as benchmark points. However, since the value of $\Lambda$ has a greater impact in the subdominant operator, we find that its effects are also negligible.

\subsection{Direct Detection Experiments} \label{subsec:DirectDetection}
\begin{figure}[h] 
\centering
\includegraphics[width=140mm]{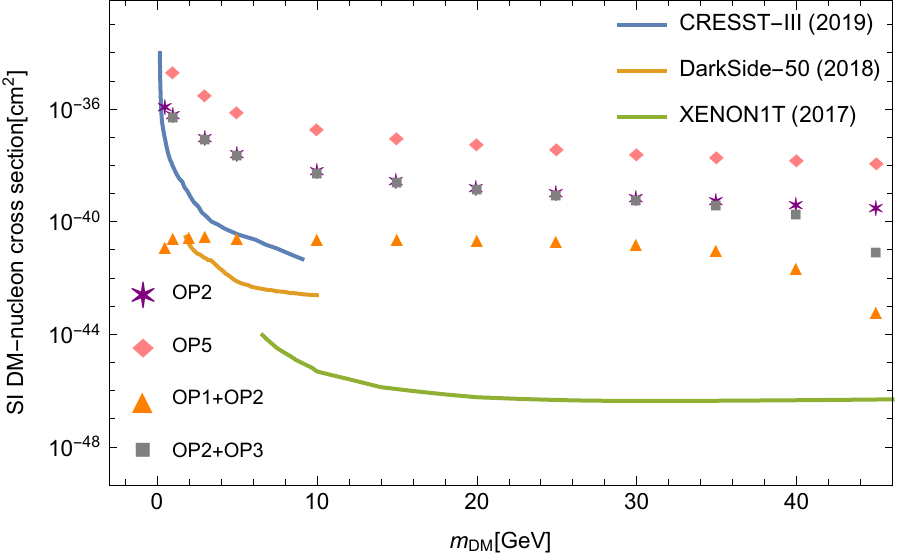}
\caption{WIMP cross sections (normalized to a single nucleon) for spin-independent coupling versus mass. OP1, OP2, OP3 and OP5 are defined in \cref{OPS}. When we combine operators, we use $\Lambda=1$ TeV. Operators not shown here have cross sections many orders of magnitude below the current limits.} \label{fig:WIMPs}
\end{figure}

Currently the most stringent limit on spin-independent scattering cross sections of DM-nucleon particles come from the XENON1T, DarkSide-50 and CRESST-III experiments. In order to derive the implications for the effective theory under study we obtained the DM-nucleon cross sections in the limit where the relative velocity goes to zero. We use micrOMEGAs \cite{micromegas} to compute it. 

Fig. \ref{fig:WIMPs} shows the values for the DM-nucleon scattering cross sections already excluded by the DarkSide-50 \cite{DarkSide},  XENON1T \cite{Aprile:2018dbl} and CRESST-III \cite{Abdelhameed:2019hmk} experiments, as well as the results that we obtained for different operators. The notation used in this figure is defined in \cref{OPS}. 

We can see that the operators included in fig. \ref{fig:WIMPs} are ruled out in the mass range we are considering, with the exception of a small region of very light masses. Operators not shown in fig. \ref{fig:WIMPs} -- OP1, OP3, OP1+OP3 and OP4 -- have DM-nucleon cross sections many orders of magnitude below the current limits from  direct detection experiments. Therefore, in the following, we will only consider those operators not shown in fig. \ref{fig:WIMPs}.

\subsection{Dwarf spheroidal satellite galaxies}\label{subsec:DSSG}
\begin{figure}[h] 
\centering
\includegraphics[width=170mm]{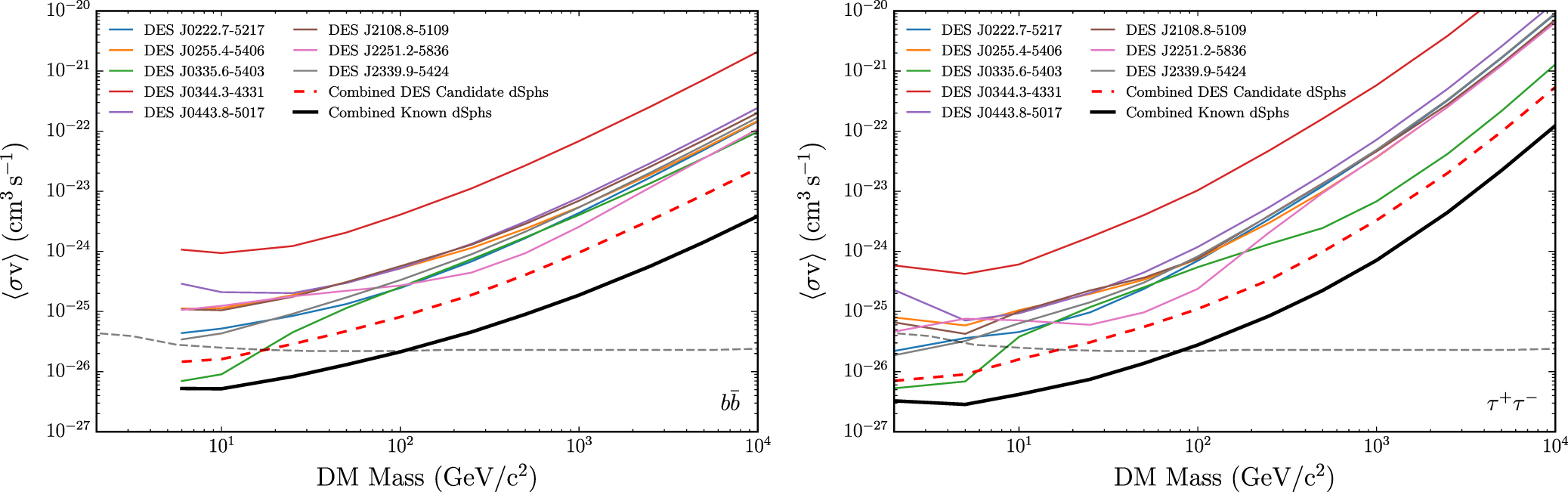}
\caption{{\small Upper limits on the velocity-averaged DM annihilation cross section at 95\%CL for DM annihilation to $\bar{b}b$(left) and $\tau^+ \tau^-$(right). The current best limits derived from a joint analysis of 15 previously known dSphs are also shown (black curve). For reference, we also display (dashed gray curve) the thermal relic cross section derived by Steigman et al. \cite{PhysRevD.86.023506}.}} \label{fig:Drlica}
\end{figure}

The dwarf spheroidal satellite galaxies (dSphs) of the Milky Way are some of the most DM dominated objects known. Due to their proximity, high DM content, and apparent absence of non-thermal processes, dSphs are excellent targets for the indirect detection of DM. Recently, eight new dSph candidates were discovered using the first year of data from the Dark Energy Survey (DES). Ref. \cite{Drlica} searched for gamma-ray emission coincident with the positions of these new objects in six years of Fermi Large Area Telescope data. No significant excesses of gamma-ray emission were found. Individual and combined limits on the velocity-averaged DM annihilation cross section for these new targets ---assuming that the DES candidates are dSphs with DM halo properties similar to the known dSphs--- were also computed, as we can see in fig. \ref{fig:Drlica}.  

\begin{figure}[H] 
\centering
\includegraphics[width=110mm]{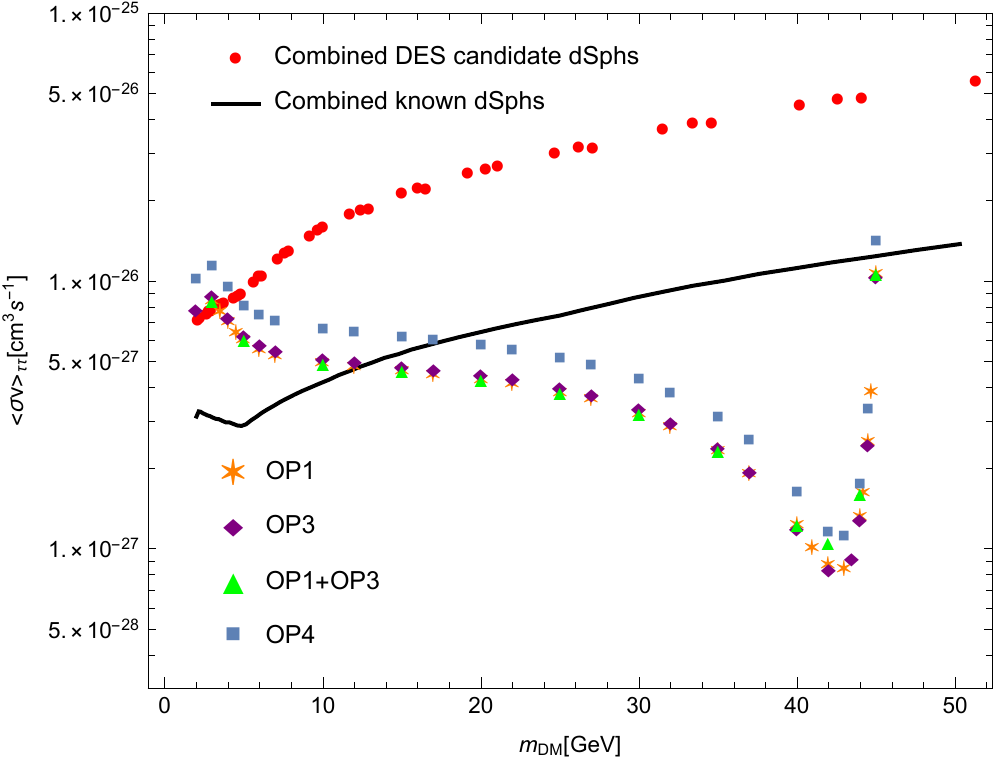}
\caption{{\small Restrictions from dSphs on the DM annihilation cross sections into $\tau^+ \tau^-$ for the portals generated by OP1, OP3 and OP4, defined in  \cref{OPS}.}} \label{fig:tau}
\end{figure}

\begin{figure}[H] 
\centering
\includegraphics[width=110mm]{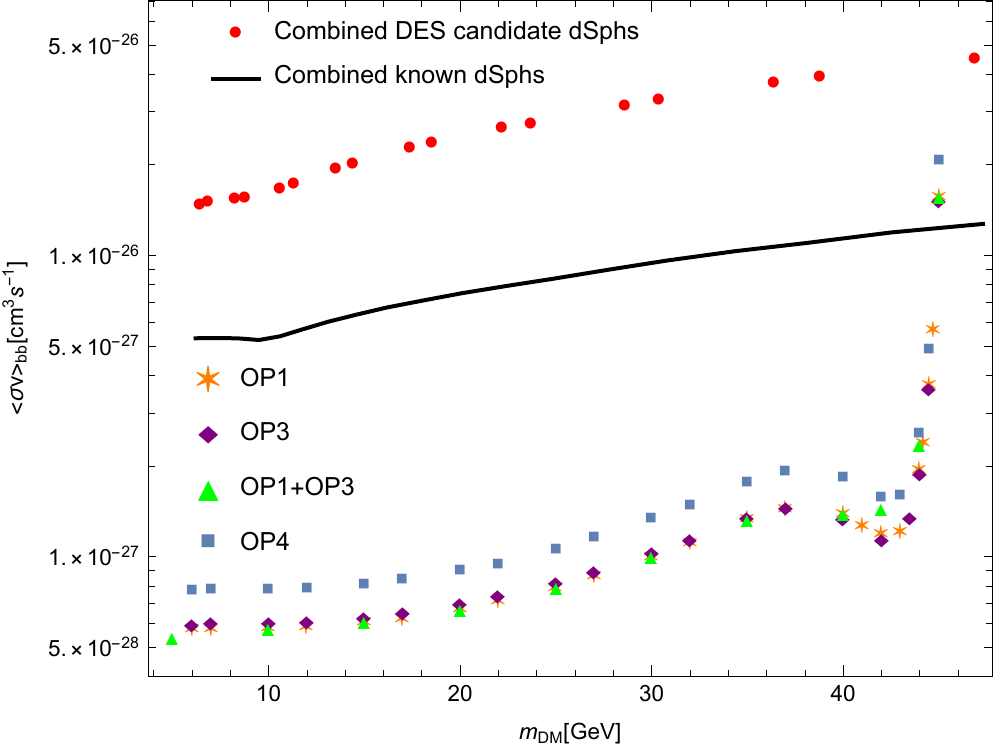}
\caption{{\small Restrictions from dSphs on the DM annihilation cross sections into $\bar{b} b$ for the portals generated by OP1, OP3 and OP4, defined in \cref{OPS}.}} \label{fig:b}
\end{figure}

We computed the non-relativistic ($m_{DM} \ll T$) thermally-averaged DM annihilation cross sections $\langle\sigma v \rangle$, using our effective operators OP1,3,4, and compared the results with the limits mentioned before. 
The results are presented in figs. \ref{fig:tau} and \ref{fig:b}, where we can see that DM masses in the intermediate region $ 10\text{ GeV} \lesssim m_{DM} \lesssim 45 \text{ GeV} $ are allowed.

\subsection{Limits from AMS-02 positron measurements}\label{subsec:AMS02}
\begin{figure}[htb] 
\centering
\includegraphics[width=150mm]{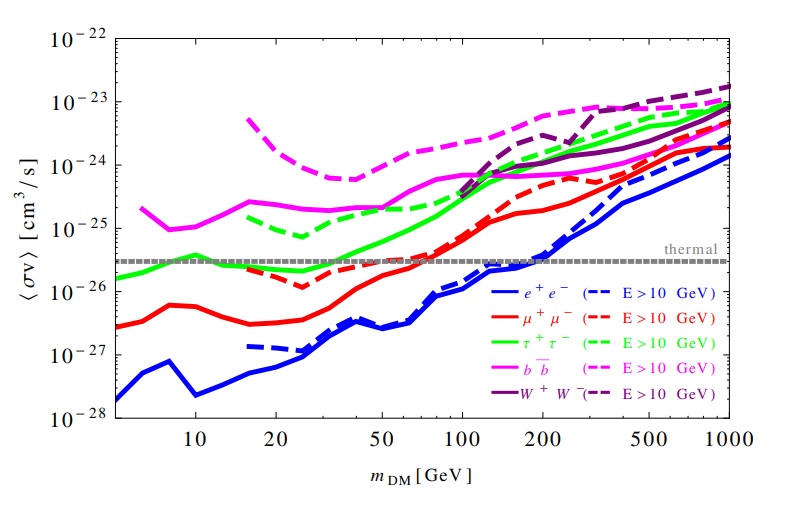}
\caption{{\footnotesize Limits on the annihilation cross section derived from the AMS-02 data on the positron fraction, assuming the MED propagation model \cite{Donato:2003xg}. The limits shown as solid lines were derived from sampling over various energy windows, while the dashed lines are from considering those windows including only data with energies above 10 GeV.}} \label{fig:AMS02}
\end{figure}

The AMS-02 Collaboration has presented high-quality measurements of positron fluxes as well as the positron fraction. Working under the well-motivated assumption that a background positron flux exists from spallations of cosmic rays with the interstellar medium and from astrophysical sources, ref. \cite{Ibarra:2013zia} used measurements of the positron flux to derive limits on the dark matter annihilation cross section and lifetime for various final states (see also ref. \cite{Bergstrom:2013jra}), and extracted strong limits on DM properties. Specifically, for DM particles annihilating only into $e^+e^-$ or into $\mu^+\mu^-$, their bounds on the annihilation cross section are stronger than the thermal value when the dark mass is in the range that we are considering. These limits are shown in fig. \ref{fig:AMS02}, where the solid lines correspond to the best limits sampling over all energy windows, while those shown with dashed lines were derived selecting windows containing only energies larger than 10 GeV. The latter limits are only mildly affected by the modeling of the solar modulation and are therefore more robust, so we will use them in the calculations below.

\begin{figure}[H] 
\centering
\includegraphics[width=110mm]{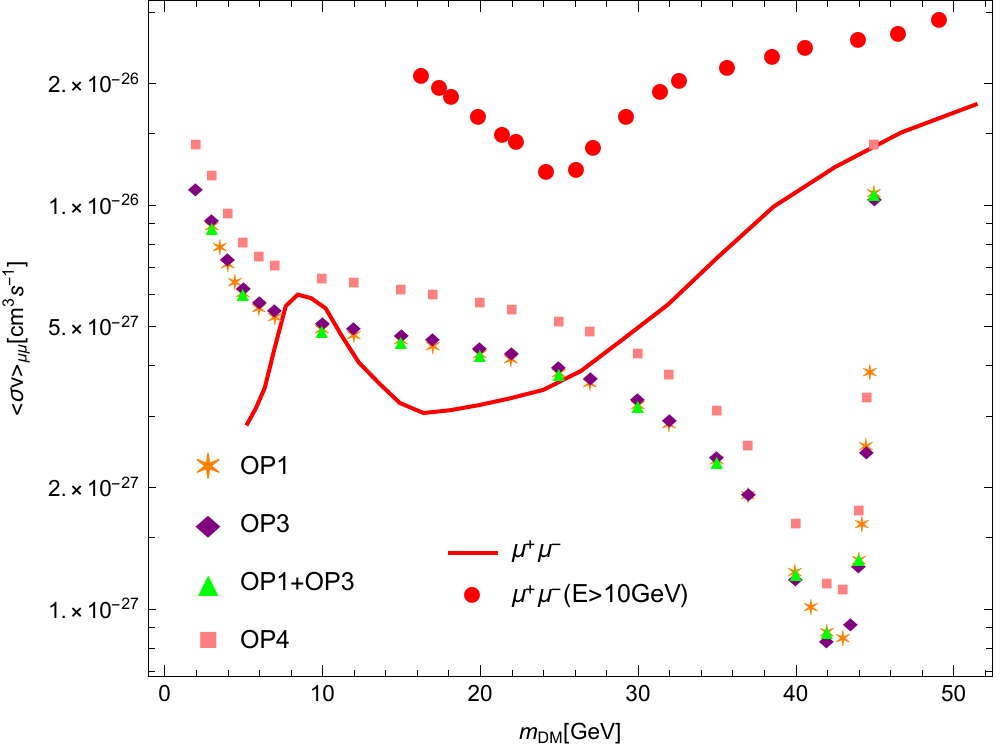}
\caption{{\small Restrictions from AMS-02 data on the DM annihilation cross sections into $\mu^+ \mu^-$ for the portals generated by OP1, OP3 and OP4, defined in \cref{OPS}.}} \label{fig:mu}
\end{figure}

\begin{figure}[H] 
\centering
\includegraphics[width=110mm]{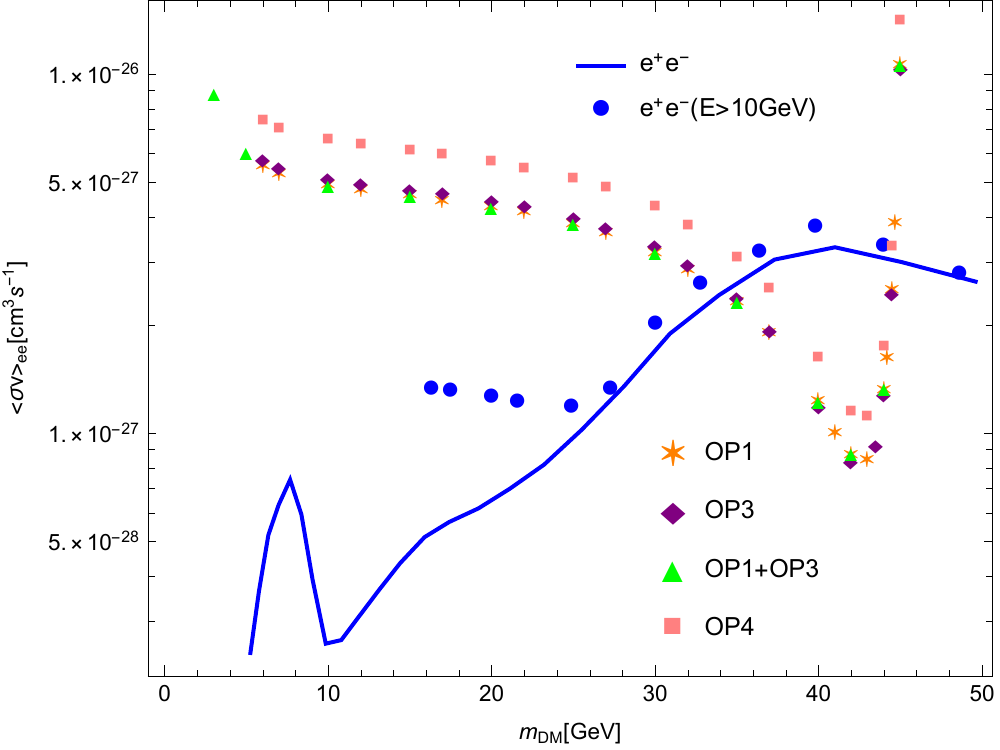}
\caption{{\small Restrictions from AMS-02 data on the DM annihilation cross sections into $e^+ e^-$ for the portals generated by OP1, OP3 and OP4, defined in \cref{OPS}.}} \label{fig:e}
\end{figure}

The comparison of the limits derived using the annihilation cross sections into $e^+e^-$ and $\mu^+ \mu^-$ final states and those computed with our effective operators are shown in figs. \ref{fig:mu} and \ref{fig:e}. Ref. \cite{Ibarra:2013zia} also derived limits for the final states $\tau^+ \tau^-,~\bar{b} b$, but these are weaker than those using the dSphs data. 

\section{Discussion and conclusions}\label{sec:Concl}

From the various limits obtained in the sections above, we derive the following restrictions on the operators listed in \cref{OPS} (items 1-5 below refer to OP1-5, respectively). The discussion below refers only to values for the effective coupling coefficients that correctly reproduce the relic density in \cref{rd}. A summary of our results\footnote{DM masses below $ 2 $ GeV are allowed for OP1 and OP4 because the updated value of $\Gamma^\text{inv}_Z - \Gamma^{\bar{\nu} \nu}_Z$  used in sect. \ref{sec:GammaZ}. The previous result (footnote \ref{ft_iv}) excluded these regions of parameter space.} is given in table \ref{tab:res}.

\begin{enumerate}

\item We obtain a negligible direct detection cross sections when we use the operator $B_{\mu \nu} \bar{\Psi} \sigma^{\mu \nu} \Psi$ (OP1), many orders of magnitude below the bounds from current or future direct detection experiments. The strongest constraint on this operator comes from thermally-averaged DM annihilation cross section into $e^+e^-$ (as we can see in fig. \ref{fig:e}) that only allows masses in the range $\approx 33-44.5$ GeV.

\item If we take the operator involving \textit{fermionic} currents in both sectors in the operator $\mathcal{J}_{\text{SM}}^\mu \mathcal{J}_{\text{dark}\,\mu}$ (OP2), we get values for the DM-nucleon cross sections that are already excluded by experiments ({\it cf.} fig. \ref{fig:WIMPs}). Therefore, we can rule out this operator for the range of masses that we are considering. 

\item When we use the operator $B_{\mu \nu} \bar{\Psi} (\gamma^\mu \overleftrightarrow{\mathcal{D}}^\nu - \gamma^\nu \overleftrightarrow{\mathcal{D}}^\mu) P_{L, R} \Psi$ (OP3), we again get a negligible direct detection cross section. The thermally-averaged DM annihilation cross section into $e^+e^-$  shown in fig. \ref{fig:e}  excludes masses  $\lesssim 33$ GeV and $\sim m_Z/2$.

\item For the operator $B_{\mu \nu} X^{\mu \nu} \Phi$ (OP4), the case $m_X \nsim m_\Phi$ is excluded by the relic abundance constraint for all values in our mass range. When $ m_\Phi \simeq m_X $ the limits from direct DM detection searches are again irrelevant, but the thermally-averaged DM annihilation cross section into $e^+e^-$ excludes masses $\lesssim 36$ GeV and $\sim m_Z/2$ (fig. \ref{fig:e}). 

\item For the operator $\mathcal{J}_{\text{SM}}^\mu \mathcal{J}_{\text{dark}\,\mu}$, with scalar dark current and fermionic SM current, (OP5), the direct detection cross sections in the mass range we consider are excluded by the current data, as show in fig. \ref{fig:WIMPs}. 

\item When we include operators OP1 and OP2 simultaneously (using the prescription of \cref{eq:combine}), the limits from DarkSide and XENON1T exclude masses above $\approx 2$ GeV \footnote{The invisible Z decay width bound \cref{Zinv} is respected for DM masses as low as $\sim2.5$ MeV for the dimension five operator, and the value of the corresponding dimensionless effective coupling still satisfies \cref{relajado} and complies with the limits below $\sim 0.2$ GeV derived in ref. \cite{Bringmann:2018cvk}.}.

\item Combining operators OP1 and OP3 (using \cref{eq:combine}), we get negligible DM-nucleon cross sections. The strongest bounds come from the thermally-averaged DM annihilation cross section into $e^+e^-$ (fig \ref{fig:e}), that excludes masses  $\lesssim 33$ GeV and $\sim m_Z/2$.

\item Finally, we also explored the possibility of having operators OP2 and OP3. In that case, all parameter space is forbidden by the results from direct detection experiments.  	
\end{enumerate}

There are stringent limits on light dark particles from the ATLAS experiment \cite{ATLAS}; however, we did not use those results because they are highly model-dependent.

\begin{table}[h!]
  \begin{center}
    \begin{tabular}{|c|c|c|c|} 
	  \hline
	  \rowcolor{lightcyan}     
      \textbf{Operator} & \textbf{Dim.} & \textbf{DM candidate} & \textbf{Allowed mass (GeV)} \\
      \hline
      1.- $B_{\mu \nu} \bar{\Psi} \sigma^{\mu \nu} \Psi$ & 5 & $\Psi$ fermion & $\approx 0.0025-2,\approx 33-44.5$ \\
      \rowcolor{cosmiclatte}      
      2.- $\bar{\psi} \gamma_\mu \psi \bar{\Psi} \gamma^\mu P_{L,R} \Psi$ & 6 & $\Psi$ fermion & none \\
      3.- {\small $B_{\mu \nu} \bar{\Psi} (\gamma^\mu \protect\olra{\mathcal{D}}^\nu - \gamma^\nu \protect\olra{\mathcal{D}}^\mu) P_{L, R} \Psi$} & 6 & $\Psi$ fermion & $\approx 33-44.5$ \\
   	  \rowcolor{cosmiclatte}      
      4.- $B_{\mu \nu} X^{\mu \nu} \Phi$ & 5 & {\small vector $X$, scalar $\Phi$} & $\approx 0.11-2, \approx 36-44.5$ \\
      5.- $\bar{\psi} \gamma_\mu \psi \frac{1}{2i} \Phi^\dagger \overleftrightarrow{\mathcal{D}}^\mu \Phi$ & 6 & scalar $\Phi$ & none \\
   	  \rowcolor{cosmiclatte}      
      $1+2$ & 5+6 & $\Psi$ fermion & $\approx 0.0025-2$ \\
      $1+3$ & 5+6 & $\Psi$ fermion & $\approx 0.0025-2, \approx 33-44.5$ \\
   	  \rowcolor{cosmiclatte}      
      $2+3$ & 6 & $\Psi$ fermion & none \\     
      \hline
    \end{tabular}
  \end{center}
  \caption{Summary of results obtained considering the $Z$ invisible decay width, relic density, direct detection experiments and indirect detection results from dSphs and positron flux measurements. It is very important to note that we are considering masses of the dark particles below the mass of the $Z$ boson ($M_Z/2\sim 45. 5$ GeV, as they appear in charge conjugated pairs). } \label{tab:res}      
\end{table}

We see then that the combination of direct detection experiments and the constraints from relic density, rule out the operators of category VII in table \ref{tab:effops}. In contrast, we found that all dimension 5 operator portals in category IV are compatible with the experimental constraints ---invisible decay width of the $Z$ boson, DM relic density\footnote{We recall the fact that we are considering the WIMP freeze-out scenario when computing the DM relic density.}, direct detection experiments, and indirect limits coming from dSphs and positron flux measurements--- for DM masses lighter than $M_Z/2$ and larger than $\sim 33$ GeV. For the allowed operator portals, the most stringent limits come from the thermally averaged DM annihilation cross section into $e^+e^-$ as shown in fig. \ref{fig:e}. Besides, we can see that these operators have an allowed region for masses between some MeV and $2$ GeV, where the upper limit comes from the DM annihilation cross sections into $\tau^+ \tau^-$ ({\it cf.} fig. \ref{fig:tau}), while the lower limits are due to the $Z$ invisible decay width. As we mentioned before, when we combine OP1+OP2 and OP1+OP3, the value of $\Lambda$ has a negligible effect, since OP1 completely dominates the interaction.

The results in table \ref{tab:res} look promising and warrant further exploration of the antisymmetric tensor, spin-1 mediator portal with mass below $m_Z$ that, together with the better-studied Higgs, fermion and vector portals may help unravel the DM puzzle, unsolved since 1933.

\section*{Acknowledgements}
This work has been partially funded by Conacyt: the support of project 250628, as well as the scholarship during F. Fortuna Ms. Sc. and Ph. D. are acknowledged. The work was also supported by the SEP-Cinvestav Fund, project number 142. \\ We would like to thank Diego Aristiz\'abal, Alexander Belyaev, Alejandro Ibarra, Alexander Pukhov and Avelino Vicente for their advice in the development of different aspects of this work.

\nocite{Belyaev:2012qa}
\nocite{Semenov:2014rea}
\bibliographystyle{unsrt}
\bibliography{bibtesis}

\begin{thebibliography}{10}

\bibitem{Zwicky:1933gu}
F.~Zwicky.
\newblock {Die Rotverschiebung von extragalaktischen Nebeln}.
\newblock {\em Helv. Phys. Acta}, 6:110--127, 1933.
\newblock [Gen. Rel. Grav.41,207(2009)].

\bibitem{Rubin:1970zza}
Vera~C. Rubin and W.~Kent Ford, Jr.
\newblock {Rotation of the Andromeda Nebula from a Spectroscopic Survey of
  Emission Regions}.
\newblock {\em Astrophys. J.}, 159:379--403, 1970.

\bibitem{Rubin:1980zd}
V.~C. Rubin, N.~Thonnard, and W.~K. Ford, Jr.
\newblock {Rotational properties of 21 SC galaxies with a large range of
  luminosities and radii, from NGC 4605 /R = 4kpc/ to UGC 2885 /R = 122 kpc/}.
\newblock {\em Astrophys. J.}, 238:471, 1980.

\bibitem{Corbelli:1999af}
Edvige Corbelli and Paolo Salucci.
\newblock {The Extended Rotation Curve and the Dark Matter Halo of M33}.
\newblock {\em Mon. Not. Roy. Astron. Soc.}, 311:441--447, 2000.

\bibitem{Clowe:2006eq}
Douglas Clowe, Marusa Bradac, Anthony~H. Gonzalez, Maxim Markevitch, Scott~W.
  Randall, Christine Jones, and Dennis Zaritsky.
\newblock {A direct empirical proof of the existence of dark matter}.
\newblock {\em Astrophys. J.}, 648:L109--L113, 2006.

\bibitem{Allen:2011zs}
Steven~W. Allen, August~E. Evrard, and Adam~B. Mantz.
\newblock {Cosmological Parameters from Observations of Galaxy Clusters}.
\newblock {\em Ann. Rev. Astron. Astrophys.}, 49:409--470, 2011.

\bibitem{Ade:2015xua}
P.~A.~R. Ade et~al.
\newblock {Planck 2015 results. XIII. Cosmological parameters}.
\newblock {\em Astron. Astrophys.}, 594:A13, 2016.

\bibitem{Fu:2016ega}
Changbo Fu et~al.
\newblock {Spin-Dependent Weakly-Interacting-Massive-Particle–Nucleon Cross
  Section Limits from First Data of PandaX-II Experiment}.
\newblock {\em Phys. Rev. Lett.}, 118(7):071301, 2017.
\newblock [Erratum: Phys. Rev. Lett.120,no.4,049902(2018)].

\bibitem{Aprile:2017iyp}
E.~Aprile et~al.
\newblock {First Dark Matter Search Results from the XENON1T Experiment}.
\newblock {\em Phys. Rev. Lett.}, 119(18):181301, 2017.

\bibitem{Akerib:2016lao}
D.~S. Akerib et~al.
\newblock {Results on the Spin-Dependent Scattering of Weakly Interacting
  Massive Particles on Nucleons from the Run 3 Data of the LUX Experiment}.
\newblock {\em Phys. Rev. Lett.}, 116(16):161302, 2016.

\bibitem{Behnke:2016lsk}
E.~Behnke et~al.
\newblock {Final Results of the PICASSO Dark Matter Search Experiment}.
\newblock {\em Astropart. Phys.}, 90:85--92, 2017.

\bibitem{Akerib:2016vxi}
D.~S. Akerib et~al.
\newblock {Results from a search for dark matter in the complete LUX exposure}.
\newblock {\em Phys. Rev. Lett.}, 118(2):021303, 2017.

\bibitem{Tan:2016zwf}
Andi Tan et~al.
\newblock {Dark Matter Results from First 98.7 Days of Data from the PandaX-II
  Experiment}.
\newblock {\em Phys. Rev. Lett.}, 117(12):121303, 2016.

\bibitem{Hanany:2019lle}
Shaul Hanany et~al.
\newblock {PICO: Probe of Inflation and Cosmic Origins}.
\newblock 2019.

\bibitem{Hooper:2010mq}
Dan Hooper and Lisa Goodenough.
\newblock {Dark Matter Annihilation in The Galactic Center As Seen by the Fermi
  Gamma Ray Space Telescope}.
\newblock {\em Phys. Lett.}, B697:412--428, 2011.

\bibitem{Bulbul:2014sua}
Esra Bulbul, Maxim Markevitch, Adam Foster, Randall~K. Smith, Michael
  Loewenstein, and Scott~W. Randall.
\newblock {Detection of An Unidentified Emission Line in the Stacked X-ray
  spectrum of Galaxy Clusters}.
\newblock {\em Astrophys. J.}, 789:13, 2014.

\bibitem{Urban:2014yda}
O.~Urban, N.~Werner, S.~W. Allen, A.~Simionescu, J.~S. Kaastra, and L.~E.
  Strigari.
\newblock {A Suzaku Search for Dark Matter Emission Lines in the X-ray
  Brightest Galaxy Clusters}.
\newblock {\em Mon. Not. Roy. Astron. Soc.}, 451(3):2447--2461, 2015.

\bibitem{Choi:2015ara}
K.~Choi et~al.
\newblock {Search for neutrinos from annihilation of captured low-mass dark
  matter particles in the Sun by Super-Kamiokande}.
\newblock {\em Phys. Rev. Lett.}, 114(14):141301, 2015.

\bibitem{Ruchayskiy:2015onc}
Oleg Ruchayskiy, Alexey Boyarsky, Dmytro Iakubovskyi, Esra Bulbul, Dominique
  Eckert, Jeroen Franse, Denys Malyshev, Maxim Markevitch, and Andrii Neronov.
\newblock {Searching for decaying dark matter in deep XMM–Newton observation
  of the Draco dwarf spheroidal}.
\newblock {\em Mon. Not. Roy. Astron. Soc.}, 460(2):1390--1398, 2016.

\bibitem{Ackermann:2015zua}
M.~Ackermann et~al.
\newblock {Searching for Dark Matter Annihilation from Milky Way Dwarf
  Spheroidal Galaxies with Six Years of Fermi Large Area Telescope Data}.
\newblock {\em Phys. Rev. Lett.}, 115(23):231301, 2015.

\bibitem{Franse:2016dln}
Jeroen Franse et~al.
\newblock {Radial Profile of the 3.55 keV line out to $R_{200}$ in the Perseus
  Cluster}.
\newblock {\em Astrophys. J.}, 829(2):124, 2016.

\bibitem{Aharonian:2016gzq}
F.~A. Aharonian et~al.
\newblock {$Hitomi$ constraints on the 3.5 keV line in the Perseus galaxy
  cluster}.
\newblock {\em Astrophys. J.}, 837(1):L15, 2017.

\bibitem{Cui:2016ppb}
Ming-Yang Cui, Qiang Yuan, Yue-Lin~Sming Tsai, and Yi-Zhong Fan.
\newblock {Possible dark matter annihilation signal in the AMS-02 antiproton
  data}.
\newblock {\em Phys. Rev. Lett.}, 118(19):191101, 2017.

\bibitem{Aartsen:2016zhm}
M.~G. Aartsen et~al.
\newblock {Search for annihilating dark matter in the Sun with 3 years of
  IceCube data}.
\newblock {\em Eur. Phys. J.}, C77(3):146, 2017.
\newblock [Erratum: Eur. Phys. J.C79,no.3,214(2019)].

\bibitem{TheFermi-LAT:2017vmf}
M.~Ackermann et~al.
\newblock {The Fermi Galactic Center GeV Excess and Implications for Dark
  Matter}.
\newblock {\em Astrophys. J.}, 840(1):43, 2017.

\bibitem{Abercrombie:2015wmb}
Daniel Abercrombie et~al.
\newblock {Dark Matter Benchmark Models for Early LHC Run-2 Searches: Report of
  the ATLAS/CMS Dark Matter Forum}.
\newblock {\em Phys. Dark Univ.}, 26:100371, 2019.

\bibitem{Aaboud:2016wna}
Morad Aaboud et~al.
\newblock {Dark matter interpretations of ATLAS searches for the electroweak
  production of supersymmetric particles in $ \sqrt{s}=8 $ TeV proton-proton
  collisions}.
\newblock {\em JHEP}, 09:175, 2016.

\bibitem{Sirunyan:2018gka}
Albert~M Sirunyan et~al.
\newblock {Search for dark matter in events with energetic, hadronically
  decaying top quarks and missing transverse momentum at $ \sqrt{s}=13 $ TeV}.
\newblock {\em JHEP}, 06:027, 2018.

\bibitem{ATLAS}
Morad Aaboud et~al.
\newblock {Constraints on mediator-based dark matter and scalar dark energy
  models using $\sqrt s = 13$ TeV $pp$ collision data collected by the ATLAS
  detector}.
\newblock {\em JHEP}, 05:142, 2019.

\bibitem{Baer:2014eja}
Howard Baer, Ki-Young Choi, Jihn~E. Kim, and Leszek Roszkowski.
\newblock {Dark matter production in the early Universe: beyond the thermal
  WIMP paradigm}.
\newblock {\em Phys. Rept.}, 555:1--60, 2015.

\bibitem{Dutra}
{Arcadi, Giorgio}, {Dutra, Ma\'{\i}ra}, {Ghosh, Pradipta}, {Lindner, Manfred},
  {Mambrini, Yann}, {Pierre, Mathias}, {Profumo, Stefano}, and {Queiroz,
  Farinaldo S.}
\newblock The waning of the wimp? a review of models, searches, and
  constraints.
\newblock {\em Eur. Phys. J. C}, 78(3):203, 2018.

\bibitem{Roszkowski:2017nbc}
Leszek Roszkowski, Enrico~Maria Sessolo, and Sebastian Trojanowski.
\newblock {WIMP dark matter candidates and searches---current status and future
  prospects}.
\newblock {\em Rept. Prog. Phys.}, 81(6):066201, 2018.

\bibitem{Belanger:2008sj}
G.~Belanger, F.~Boudjema, A.~Pukhov, and A.~Semenov.
\newblock {Dark matter direct detection rate in a generic model with micrOMEGAs
  2.2}.
\newblock {\em Comput. Phys. Commun.}, 180:747--767, 2009.

\bibitem{Goodman:2010ku}
Jessica Goodman, Masahiro Ibe, Arvind Rajaraman, William Shepherd, Tim M.~P.
  Tait, and Hai-Bo Yu.
\newblock {Constraints on Dark Matter from Colliders}.
\newblock {\em Phys. Rev.}, D82:116010, 2010.

\bibitem{Crivellin:2014gpa}
Andreas Crivellin and Ulrich Haisch.
\newblock {Dark matter direct detection constraints from gauge bosons loops}.
\newblock {\em Phys. Rev.}, D90:115011, 2014.

\bibitem{Crivellin:2014qxa}
Andreas Crivellin, Francesco D'Eramo, and Massimiliano Procura.
\newblock {New Constraints on Dark Matter Effective Theories from Standard
  Model Loops}.
\newblock {\em Phys. Rev. Lett.}, 112:191304, 2014.

\bibitem{Duch:2014xda}
Mateusz Duch, Bohdan Grzadkowski, and Jose Wudka.
\newblock {Classification of effective operators for interactions between the
  Standard Model and dark matter}.
\newblock {\em JHEP}, 05:116, 2015.

\bibitem{Wudka}
Vannia Mac\'ias-Gonz\'alez and Jos\'e Wudka.
\newblock Effective theories for dark matter interactions and the neutrino
  portal paradigm.
\newblock {\em Journal of High Energy Physics}, 07 2015.

\bibitem{Arcadi:2019lka}
Giorgio Arcadi, Abdelhak Djouadi, and Martti Raidal.
\newblock {Dark Matter through the Higgs portal}.
\newblock 2019.

\bibitem{Cosme:2005sb}
Nicolas Cosme, Laura Lopez~Honorez, and Michel H.~G. Tytgat.
\newblock {Leptogenesis and dark matter related?}
\newblock {\em Phys. Rev.}, D72:043505, 2005.

\bibitem{An:2009vq}
Haipeng An, Shao-Long Chen, Rabindra~N. Mohapatra, and Yue Zhang.
\newblock {Leptogenesis as a Common Origin for Matter and Dark Matter}.
\newblock {\em JHEP}, 03:124, 2010.

\bibitem{Falkowski:2009yz}
Adam Falkowski, Jose Juknevich, and Jessie Shelton.
\newblock {Dark Matter Through the Neutrino Portal}.
\newblock 2009.

\bibitem{Lindner:2010rr}
Manfred Lindner, Alexander Merle, and Viviana Niro.
\newblock {Enhancing Dark Matter Annihilation into Neutrinos}.
\newblock {\em Phys. Rev.}, D82:123529, 2010.

\bibitem{Farzan:2011ck}
Yasaman Farzan.
\newblock {Flavoring Monochromatic Neutrino Flux from Dark Matter
  Annihilation}.
\newblock {\em JHEP}, 02:091, 2012.

\bibitem{Falkowski:2011xh}
Adam Falkowski, Joshua~T. Ruderman, and Tomer Volansky.
\newblock {Asymmetric Dark Matter from Leptogenesis}.
\newblock {\em JHEP}, 05:106, 2011.

\bibitem{Heeck:2012bz}
Julian Heeck and He~Zhang.
\newblock {Exotic Charges, Multicomponent Dark Matter and Light Sterile
  Neutrinos}.
\newblock {\em JHEP}, 05:164, 2013.

\bibitem{Baek:2013qwa}
Seungwon Baek, P.~Ko, and Wan-Il Park.
\newblock {Singlet Portal Extensions of the Standard Seesaw Models to a Dark
  Sector with Local Dark Symmetry}.
\newblock {\em JHEP}, 07:013, 2013.

\bibitem{Baldes:2015lka}
Iason Baldes, Nicole~F. Bell, Alexander~J. Millar, and Raymond~R. Volkas.
\newblock {Asymmetric Dark Matter and CP Violating Scatterings in a UV Complete
  Model}.
\newblock {\em JCAP}, 1510:048, 2015.

\bibitem{VaniaIllana}
Vannia González-Macías, José~I. Illana, and José Wudka.
\newblock {A realistic model for Dark Matter interactions in the neutrino
  portal paradigm}.
\newblock {\em JHEP}, 05:171, 2016.

\bibitem{Batell:2017rol}
Brian Batell, Tao Han, and Barmak Shams Es~Haghi.
\newblock {Indirect Detection of Neutrino Portal Dark Matter}.
\newblock {\em Phys. Rev.}, D97(9):095020, 2018.

\bibitem{HajiSadeghi:2017zrl}
S.~HajiSadeghi, S.~Smolenski, and J.~Wudka.
\newblock {Asymmetric dark matter with a possible Bose-Einstein condensate}.
\newblock {\em Phys. Rev.}, D99(2):023514, 2019.

\bibitem{Bandyopadhyay:2018qcv}
Priyotosh Bandyopadhyay, Eung~Jin Chun, Rusa Mandal, and Farinaldo~S. Queiroz.
\newblock {Scrutinizing Right-Handed Neutrino Portal Dark Matter With Yukawa
  Effect}.
\newblock {\em Phys. Lett.}, B788:530--534, 2019.

\bibitem{Berlin:2018ztp}
Asher Berlin and Nikita Blinov.
\newblock {Thermal neutrino portal to sub-MeV dark matter}.
\newblock {\em Phys. Rev.}, D99(9):095030, 2019.

\bibitem{Blennow:2019fhy}
M.~Blennow, E.~Fernandez-Martinez, A.~Olivares-Del~Campo, S.~Pascoli,
  S.~Rosauro-Alcaraz, and A.~V. Titov.
\newblock {Neutrino Portals to Dark Matter}.
\newblock {\em Eur. Phys. J.}, C79(7):555, 2019.

\bibitem{pdg}
M.~Tanabashi et~al. (Particle Data~Group).
\newblock {\em Phys. Rev. D}, 98(030001), 2018.

\bibitem{Aprile:2018dbl}
E.~Aprile et~al.
\newblock {Dark Matter Search Results from a One Ton-Year Exposure of XENON1T}.
\newblock {\em Phys. Rev. Lett.}, 121(11):111302, 2018.

\bibitem{Ren:2018gyx}
Xiangxiang Ren et~al.
\newblock {Constraining Dark Matter Models with a Light Mediator at the
  PandaX-II Experiment}.
\newblock {\em Phys. Rev. Lett.}, 121(2):021304, 2018.

\bibitem{Akerib:2018hck}
D.~S. Akerib et~al.
\newblock {Results of a Search for Sub-GeV Dark Matter Using 2013 LUX Data}.
\newblock {\em Phys. Rev. Lett.}, 122(13):131301, 2019.

\bibitem{Agnes:2018oej}
P.~Agnes et~al.
\newblock {Constraints on Sub-GeV Dark-Matter–Electron Scattering from the
  DarkSide-50 Experiment}.
\newblock {\em Phys. Rev. Lett.}, 121(11):111303, 2018.

\bibitem{Abdelhameed:2019hmk}
A.~H. Abdelhameed et~al.
\newblock {First results from the CRESST-III low-mass dark matter program}.
\newblock {\em Phys. Rev.}, D100(10):102002, 2019.

\bibitem{Drlica}
A.~Drlica-Wagner et~al.
\newblock {Search for Gamma-Ray Emission from DES Dwarf Spheroidal Galaxy
  Candidates with Fermi-LAT Data}.
\newblock {\em Astrophys. J.}, 809(1):L4, 2015.

\bibitem{Ibarra:2013zia}
fndro Ibarra, Anna~S. Lamperstorfer, and Joseph Silk.
\newblock {Dark matter annihilations and decays after the AMS-02 positron
  measurements}.
\newblock {\em Phys. Rev.}, D89(6):063539, 2014.

\bibitem{Accardo:2014lma}
L.~Accardo et~al.
\newblock {High Statistics Measurement of the Positron Fraction in Primary
  Cosmic Rays of 0.5–500 GeV with the Alpha Magnetic Spectrometer on the
  International Space Station}.
\newblock {\em Phys. Rev. Lett.}, 113:121101, 2014.

\bibitem{darkmatter}
Dan~Hooper Gianfranco~Bertone and Joseph Silk.
\newblock Particle dark matter: evidence, candidates and constrains.
\newblock {\em Physics Reports}, 405(5-6):279--390, January 2005.

\bibitem{Feng:2010gw}
Jonathan~L. Feng.
\newblock {Dark Matter Candidates from Particle Physics and Methods of
  Detection}.
\newblock {\em Ann. Rev. Astron. Astrophys.}, 48:495--545, 2010.

\bibitem{Racco:2015dxa}
Davide Racco, Andrea Wulzer, and Fabio Zwirner.
\newblock {Robust collider limits on heavy-mediator Dark Matter}.
\newblock {\em JHEP}, 05:009, 2015.

\bibitem{Bell:2015sza}
Nicole~F. Bell, Yi~Cai, James~B. Dent, Rebecca~K. Leane, and Thomas~J. Weiler.
\newblock {Dark matter at the LHC: Effective field theories and gauge
  invariance}.
\newblock {\em Phys. Rev.}, D92(5):053008, 2015.

\bibitem{DeSimone:2016fbz}
Andrea De~Simone and Thomas Jacques.
\newblock {Simplified models vs. effective field theory approaches in dark
  matter searches}.
\newblock {\em Eur. Phys. J.}, C76(7):367, 2016.

\bibitem{Cao:2009uw}
Qing-Hong Cao, Chuan-Ren Chen, Chong~Sheng Li, and Hao Zhang.
\newblock {Effective Dark Matter Model: Relic density, CDMS II, Fermi LAT and
  LHC}.
\newblock {\em JHEP}, 08:018, 2011.

\bibitem{Cheung:2012gi}
Kingman Cheung, Po-Yan Tseng, Yue-Lin~S. Tsai, and Tzu-Chiang Yuan.
\newblock {Global Constraints on Effective Dark Matter Interactions: Relic
  Density, Direct Detection, Indirect Detection, and Collider}.
\newblock {\em JCAP}, 1205:001, 2012.

\bibitem{Busoni:2013lha}
Giorgio Busoni, Andrea De~Simone, Enrico Morgante, and Antonio Riotto.
\newblock {On the Validity of the Effective Field Theory for Dark Matter
  Searches at the LHC}.
\newblock {\em Phys. Lett.}, B728:412--421, 2014.

\bibitem{Buchmueller:2014yoa}
Oliver Buchmueller, Matthew~J. Dolan, Sarah~A. Malik, and Christopher McCabe.
\newblock {Characterising dark matter searches at colliders and direct
  detection experiments: Vector mediators}.
\newblock {\em JHEP}, 01:037, 2015.

\bibitem{Patt:2006fw}
Brian Patt and Frank Wilczek.
\newblock {Higgs-field portal into hidden sectors}.
\newblock 2006.

\bibitem{Lamprea19}
J.~M. Lamprea, E.~Peinado, S.~Smolenski, and J.~Wudka.
\newblock {Strongly Interacting Neutrino Portal Dark Matter}.
\newblock 2019.

\bibitem{Okun:1982xi}
L.B. Okun.
\newblock {LIMITS OF ELECTRODYNAMICS: PARAPHOTONS?}
\newblock {\em Sov. Phys. JETP}, 56:502, 1982.

\bibitem{Holdom:1985ag}
Bob Holdom.
\newblock {Two U(1)'s and Epsilon Charge Shifts}.
\newblock {\em Phys. Lett. B}, 166:196--198, 1986.

\bibitem{Baek:2011aa}
Seungwon Baek, P.~Ko, and Wan-Il Park.
\newblock {Search for the Higgs portal to a singlet fermionic dark matter at
  the LHC}.
\newblock {\em JHEP}, 02:047, 2012.

\bibitem{LopezHonorez:2012kv}
Laura Lopez-Honorez, Thomas Schwetz, and Jure Zupan.
\newblock {Higgs portal, fermionic dark matter, and a Standard Model like Higgs
  at 125 GeV}.
\newblock {\em Phys. Lett. B}, 716:179--185, 2012.

\bibitem{Berlin:2015wwa}
Asher Berlin, Stefania Gori, Tongyan Lin, and Lian-Tao Wang.
\newblock {Pseudoscalar Portal Dark Matter}.
\newblock {\em Phys. Rev. D}, 92:015005, 2015.

\bibitem{Bai:2013iqa}
Yang Bai and Joshua Berger.
\newblock {Fermion Portal Dark Matter}.
\newblock {\em JHEP}, 11:171, 2013.

\bibitem{Klasen:2016qux}
Michael Klasen, Florian Lyonnet, and Farinaldo~S. Queiroz.
\newblock {NLO+NLL collider bounds, Dirac fermion and scalar dark matter in the
  B–L model}.
\newblock {\em Eur. Phys. J.}, C77(5):348, 2017.

\bibitem{Kalb:1974yc}
Michael Kalb and Pierre Ramond.
\newblock {Classical direct interstring action}.
\newblock {\em Phys. Rev.}, D9:2273--2284, 1974.

\bibitem{Rohm:1985jv}
R.~Rohm and Edward Witten.
\newblock {The Antisymmetric Tensor Field in Superstring Theory}.
\newblock {\em Annals Phys.}, 170:454, 1986.

\bibitem{Cata:2014sta}
Oscar Cata and Alejandro Ibarra.
\newblock {Dark Matter Stability without New Symmetries}.
\newblock {\em Phys. Rev. D}, 90(6):063509, 2014.

\bibitem{Janot:2019oyi}
Patrick Janot and Stanisław Jadach.
\newblock {Improved Bhabha cross section at LEP and the number of light
  neutrino species}.
\newblock 2019.

\bibitem{micromegas}
G.~Bélanger, F.~Boudjema, A.~Pukhov, and A.~Semenov.
\newblock {micrOMEGAs4.1: two dark matter candidates}.
\newblock {\em Comput. Phys. Commun.}, 192:322--329, 2015.

\bibitem{pdg19}
M.~Tanabashi et~al. (Particle Data~Group).
\newblock {\em Phys. Rev. D}, 98(030001), 2019.

\bibitem{DarkSide}
P.~Agnes et~al.
\newblock {Low-Mass Dark Matter Search with the DarkSide-50 Experiment}.
\newblock {\em Phys. Rev. Lett.}, 121(8):081307, 2018.

\bibitem{PhysRevD.86.023506}
Gary Steigman, Basudeb Dasgupta, and John~F. Beacom.
\newblock Precise relic wimp abundance and its impact on searches for dark
  matter annihilation.
\newblock {\em Phys. Rev. D}, 86:023506, Jul 2012.

\bibitem{Bergstrom:2013jra}
Lars Bergstrom, Torsten Bringmann, Ilias Cholis, Dan Hooper, and Christoph
  Weniger.
\newblock {New Limits on Dark Matter Annihilation from AMS Cosmic Ray Positron
  Data}.
\newblock {\em Phys. Rev. Lett.}, 111:171101, 2013.

\bibitem{Donato:2003xg}
F.~Donato, Nicolao Fornengo, D.~Maurin, and P.~Salati.
\newblock {Antiprotons in cosmic rays from neutralino annihilation}.
\newblock {\em Phys. Rev.}, D69:063501, 2004.

\bibitem{Bringmann:2018cvk}
Torsten Bringmann and Maxim Pospelov.
\newblock {Novel direct detection constraints on light dark matter}.
\newblock {\em Phys. Rev. Lett.}, 122(17):171801, 2019.

\bibitem{Belyaev:2012qa}
Alexander Belyaev, Neil~D. Christensen, and Alexander Pukhov.
\newblock {CalcHEP 3.4 for collider physics within and beyond the Standard
  Model}.
\newblock {\em Comput. Phys. Commun.}, 184:1729--1769, 2013.

\bibitem{Semenov:2014rea}
A.~Semenov.
\newblock {LanHEP — A package for automatic generation of Feynman rules from
  the Lagrangian. Version 3.2}.
\newblock {\em Comput. Phys. Commun.}, 201:167--170, 2016.

\end{thebibliography}

\end{document}